\documentclass[aps,physrev,preprint,superscriptaddress
,nofootinbib]{revtex4-2}

\usepackage{here}
\usepackage{amsfonts}
\usepackage{amsmath,amssymb}
\usepackage{bm}
\usepackage[pdftex]{graphicx}          
\usepackage{ascmac}
\usepackage{xcolor}
\usepackage{braket}
\usepackage{here}
\usepackage{afterpage}
\usepackage{cases}
\usepackage{fancybox}
\usepackage{youngtab}
\usepackage{enumitem}
\usepackage{ulem}
\usepackage[colorlinks=true,linkcolor=blue,citecolor=blue,urlcolor=blue]{hyperref}

\begin{document}


\newcommand{\Ab}{\bar{A}}
\newcommand{\Fb}{\bar{F}}
\renewcommand{\rm}[1]{\mathrm{#1}}
\def\Tr{\qopname\relax o{Tr}}
\newcommand{\cA}{{\mathcal A}}
\newcommand{\cB}{{\mathcal B}}
\newcommand{\cC}{{\mathcal C}}
\newcommand{\cD}{{\mathcal D}}
\newcommand{\cE}{{\mathcal E}}
\newcommand{\cF}{{\mathcal F}}
\newcommand{\cH}{{\mathcal H}}
\newcommand{\cI}{{\mathcal I}}
\newcommand{\cK}{{\mathcal K}}
\newcommand{\cL}{{\mathcal L}}
\newcommand{\cM}{{\mathcal M}}
\newcommand{\cN}{{\mathcal N}}
\newcommand{\cO}{{\mathcal O}}
\newcommand{\cP}{{\mathcal P}}
\newcommand{\cQ}{{\mathcal Q}}
\newcommand{\cR}{{\mathcal R}}
\newcommand{\cS}{{\mathcal S}}
\newcommand{\cT}{{\mathcal T}}
\newcommand{\cV}{{\mathcal V}}
\newcommand{\cY}{{\mathcal Y}}
\newcommand{\cZ}{{\mathcal Z}}
\newcommand{\bC}{{\mathbb C}}
\newcommand{\bR}{{\mathbb R}}
\newcommand{\bS}{{\mathbb S}}
\newcommand{\bZ}{{\mathbb Z}}
\newcommand{\bGam}{{\boldsymbol \Gamma}}
\newcommand{\rx}{{\mathrm x}}
\newcommand{\sfa}{{\mathsf a}}
\newcommand{\sfh}{{\mathsf h}}
\newcommand{\sfw}{{\mathsf w}}
\newcommand{\Sym}{S_{\rm YM}}
\newcommand{\Scs}{S_{\rm CS}}
\newcommand{\Sir}{S_{\rm IR}}
\newcommand{\zuv}{z_{\rm UV}}
\newcommand{\zir}{z_{\rm{IR}}}
\newcommand{\tr}{{\rm tr}}
\newcommand{\wh}{\widehat}
\newcommand{\wt}{\widetilde}
\newcommand{\ha}{\hat{a}}
\newcommand{\hA}{\hat{A}}
\newcommand{\hF}{\hat{F}}
\newcommand{\hL}{\hat{L}}
\newcommand{\hR}{\hat{R}}
\newcommand{\hS}{\hat{S}}
\newcommand{\hPi}{\hat{\Pi}}
\newcommand{\dq}{\dot{q}}
\newcommand{\nn}{\nonumber}
\renewcommand{\ol}{\overline}  
\newcommand{\MeV}{\mathrm{MeV}}


\title{Dense matter in a holographic hard-wall model of QCD}


\author{Daisuke Fujii}
\affiliation{Advanced Science Research Center, Japan Atomic Energy Agency (JAEA), Tokai, 319-1195, Japan}
\affiliation{Research Center for Nuclear Physics (RCNP), Osaka University, Ibaraki 567-0048, Japan}

\author{Atsushi Hosaka}
\affiliation{Advanced Science Research Center, Japan Atomic Energy Agency (JAEA), Tokai, 319-1195, Japan}
\affiliation{Research Center for Nuclear Physics (RCNP), Osaka University, Ibaraki 567-0048, Japan}

\author{Akihiro Iwanaka}
\affiliation{Yukawa Institute for Theoretical Physics, Kyoto University, Kyoto 606-8502, Japan}
\affiliation{Research Center for Nuclear Physics (RCNP), Osaka University, Ibaraki 567-0048, Japan}

\author{Tadakatsu Sakai}
\affiliation{Kobayashi-Maskawa Institute for the Origin of Particles 
and the Universe, Nagoya University, Nagoya 464-8602, Japan}
\affiliation{Department of Physics, Nagoya University, Nagoya 464-8602, Japan}
\affiliation{KEK Theory Center, Institute of Particle and Nuclear Studies, High Energy Accelerator Research Organization, 1-1 Oho, Tsukuba, Ibaraki 305-0801, Japan}

\author{Motoi Tachibana}
\affiliation{Department of Physics, Saga University, Saga 840-8502, Japan}
\affiliation{Center for Theoretical Physics, Khazar University, 41 Mehseti Street, Baku, AZ1096, Azerbaijan}


\date{\today}

\begin{abstract}
A deeper understanding of QCD matter at strong coupling remains challenging due to its nonperturbative nature. 
To this end, we study a two-flavor holographic hard-wall model to investigate the properties of QCD at finite-density and zero temperature with a nonvanishing quark mass. 
A dense matter phase is described by a classical solution of the equations of motion in a homogeneous ansatz. 
We apply holographic renormalization to formulate the holographic dictionary that relates UV boundary data in the bulk with the physical quantities in QCD. 
We emphasize a role played by an IR boundary action on the hard wall when analyzing the QCD phase structures in this holographic setup. 
It is found that a baryonic matter phase is manifested in this model with a high baryon number density and a nearly vanishing chiral condensate. 
We derive the equation of state for the resulting phase and use it to work out the mass-radius relation for neutron stars. 
We find that the maximum mass of neutron stars can exceed two solar masses for a wide range of free parameters in this model. 
We also comment on an alternative scenario about the phase structure such that the baryonic matter phase arises at a baryon number chemical potential greater than a critical value.
\end{abstract}


\maketitle

\section{Introduction}

A significant goal of physics of modern quantum chromodynamics (QCD) is to establish the phase diagram of QCD.
The finite-temperature region with zero or low density in the QCD phase diagram is well determined from the lattice simulations~\cite{Aoki:2006we,Fodor:2001pe}.
In addition, the perturbative QCD approach determines the extremely high-temperature or high-density aspects of QCD.
The properties in the other region are still unclear due to the strong coupling and nonlinear properties of low-energy QCD.

An interesting part of the QCD phase diagram is the low-temperature and high-density region around $2-10$ times the normal nuclear density.
Understanding the physics in the region is an open problem.
Various theoretical studies propose exotic phases, typical examples being a pion or kaon condensed phase and a quarkyonic phase \cite{Migdal:1978az,Kaplan:1986yq,McLerran:2007qj}.
In addition, recent astronomical observations inform us about the matter governed by strong coupling QCD inside the neutron stars.
The realization of a high-density QCD matter is expected inside cores of neutron stars.
In particular, the discovery of a neutron star with a mass about twice the mass of the sun imposes a strong constraint on equations of state of the high-density QCD matter~\cite{Demorest:2010bx}.

Meanwhile, the investigation for the high-density matter is quite challenging:
the lattice Monte Carlo simulation has a technical problem known as the sign problem in the region.
Alternative methods for numerical calculations have been proposed but these are still at the stage of attempt today.
Furthermore, the perturbative QCD has no sign problem, but it does not help because
the coupling constant in the matter is considered to be strong~\cite{Baym:2019iky}.

An available tool to investigate the low-temperature QCD matter is an effective model calculation of QCD.
Among various models, in this work we consider an approach of holographic QCD.
It is an effective method of QCD at strong coupling based on holographic correspondence.
The holographic correspondence claims the quantum field theory(QFT) at strong coupling in $d$-dimensional spacetime is dual to a $(d+1)$-dimensional gravity theory at weak coupling.
One of the most successful realizations of the holographic correspondence is the AdS/CFT correspondence, which relates the $d$-dimensional conformal field theory (CFT) to the $(d+1)$-dimensional anti--de Sitter (AdS) spacetime.
The typical and original example of the AdS/CFT correspondence is a relation between four-dimensional $\mathcal{N}=4$ super Yang-Mills theory in the large $N_c$ limit and a five-dimensional supergravity theory in AdS spacetime~\cite{Maldacena:1997re}.
The correspondence is extended to Yang-Mills theory and QCD in the confined phase~\cite{Witten:1998zw,Sakai:2004cn,Sakai:2005yt}.
The holographic frameworks based on string theory are called a top-down model.
The consistency of string theory sometimes makes it intractable to compute 
physical quantities of interest.

The other class of holographic models is a bottom-up holographic model.
While it is an effective model inspired by the top-down holographic model, it is
constructed in terms of symmetries and conserved currents, conformal dimensions of QFT operators, and parameters in dual-QFT theories.
It is unclear whether the bottom-up models allow a string theory realization,
but the model is useful for investigating a wide range of phenomena.
A simple bottom-up holographic model is a hard-wall model proposed {in Ref.}~\cite{Erlich:2005qh,DaRold:2005mxj}.
We regard it as a holographic dual of a confined QCD-like theory at strong coupling.

There are several studies of neutron stars using holography~\cite{deBoer:2009wk,Hoyos:2016zke,BitaghsirFadafan:2020otb,Ghoroku:2021fos,Kovensky:2021kzl,Ghoroku:2023uxr}.
Recently, {Ref.~\cite{Bartolini:2022rkl} studied a phase structure of QCD for finite temperature and density by using the hard-wall model with massless quarks with the high-density matter.
It is described by the bulk gauge fields in a homogeneous ansatz, which is first proposed in~\cite{Rozali:2007rx}.}
Reference~\cite{Bartolini:2022rkl} also studies the equations of state of the high-density matter, which are used to discuss mass-radius relations of neutron stars.

In this paper, we employ the hard-wall model with two-flavor massive quarks.
Baryonic matter is described as a homogeneous matter following Ref.~\cite{Bartolini:2022rkl}.
Several thermodynamic quantities and expectation values are computed using the Gubser-Klebanov-Polyakov-Witten (GKP-W) dictionary~\cite{Gubser:1998bc,Witten:1998qj}, which is obtained by performing holographic renormalization.
We propose that each QCD phase is characterized by a choice of IR boundary conditions.
We derive the equation of state for the high-density matter, and study the mass-radius relation of neutron stars by solving the Tolman-Oppenheimer-Volkov (TOV)~\cite{Oppenheimer:1939ne} equation together with the equation of state.

This present paper is organized as follows.
In Sec. II, our theoretical framework is introduced.
Section III demonstrates
some numerical results obtained by using the results in Sec. II.
In Sec IV, we discuss physical interpretations, validities of our results, and outlooks of our study.
{In the Appendix, we give a brief review of the Hamilton-Jacobi equation, which is necessary to examine the QCD phase structure in the holographic model.}

\section{Homogeneous ansatz}

Let us start with the following action:
\begin{align}
	S &= S_g+S_{\rm{CS}}+ S_{\Phi},\\\label{Sgauge}
	S_{g}&= -\frac{M_5}{2}\int d^4xdz\; a(z)\left[\Tr\left[L_{MN}L^{MN} \right]+\frac{1}{2}\widehat{L}_{MN}\widehat{L}^{MN} + \left\{R\leftrightarrow L\right\}\right],\\\label{SCS}
	S_{\rm{CS}}&= \frac{N_c}{16\pi^2}\int d^4xdz\;\frac{1}{4}\epsilon^{MNOPQ}\left[\widehat{L}_M\left\{\Tr\left[L_{NO}L_{PQ}\right]+\frac{1}{6}\widehat{L}_{NO}\widehat{L}_{PQ}\right\} -\left\{R\leftrightarrow L\right\}\right],\\
	S_{\Phi}&= M_5 \int d^4xdz\; a^3(z)\left\{\Tr\left[\left(D_M\Phi\right)^\dagger D^M\Phi\right] -a^2(z)M^2_{\Phi}\Tr \left[\Phi^\dagger\Phi\right]
	{
		+\frac{g}{2}\mathrm{Tr}[(\Phi^\dag\Phi)^2]
	}
	\right\}.
\end{align}
This theory is defined in a cut-off AdS$_5$
in the Poincar\'{e} patch with the metric given by
\begin{align}
  ds^2
=a^2(z)\eta_{MN}dx^Mdx^N
=a^2(z)\left(\eta_{\mu\nu}dx^\mu x^\nu-dz^2 \right) \ ,
~~~
a(z) = \frac{1}{z} \ ,\nn
\end{align}
with $0\le z\le z_{\rm{IR}}$.
$\eta_{\mu\nu}$ is the 4D flat Minkowski metric.
The 5D spacetime indices $M,N,O,P,$ and $Q$ are contracted with
the 5D Minkowski metric $\eta_{MN}$.
The parameters $M_5$ and $M_\Phi^2$ are taken as \cite{Erlich:2005qh}
\begin{align}
M_5=\frac{N_c}{12\pi^2} \ ,~~~
M_{\Phi}^2=-3 \ .
\end{align}
In this paper, we work in {units} of the AdS radius $L=1$. 
The $L$ dependence of all the physical quantities can be 
recovered by dimensional analysis.
The gauge symmetry is $U(2)\times U(2)$ with the gauge
field given by $\cL_M$ and $\cR_M$.
These are decomposed into the $SU(2)\times U(1)$ gauge fields as
\begin{align}
\mathcal{L}_M = L_M^a \frac{\tau^a}{2} + \widehat{L}_M\frac{I_2}{2}\ ,
~~~
\mathcal{R}_M = R_M^a \frac{\tau^a}{2} + \widehat{R}_M\frac{I_2}{2}\ ,
\end{align}
where $\tau^a\ (a=1,2,3)$ are the Pauli matrices and $I_2$ the $2\times 2$ 
unit matrix.	
The field strength is given by
\begin{align}
  \mathcal{L}_{MN} = \partial_{M}\mathcal{L}_{N} - \partial_{N}\mathcal{L}_{M} -i\left[\mathcal{L}_M,\mathcal{L}_N\right] \ ,~~~
  \mathcal{R}_{MN} = \partial_{M}\mathcal{R}_{N} - \partial_{N}\mathcal{R}_{M} -i\left[\mathcal{R}_M,\mathcal{R}_N\right] \ .
\end{align}
$D_M\Phi$ is the covariant derivative defined by
\begin{align}
	D_M\Phi = \partial_M \Phi -i\mathcal{L}_M\Phi +i \Phi \mathcal{R}_M.
\end{align}

The homogeneous ansatz~\cite{Rozali:2007rx} is given by
\begin{align}
  L_i&=-R_i=-H(z)\frac{\tau^i}{2} \ ,\nn
\\
\hL_0&=\hR_0=\ha_0(z) \ ,
\label{homo:ansatz}\\
\Phi&=\frac{I_2}{2}\omega_0(z) \nn
\end{align}
with gauge fixing conditions
\begin{align}
	\hat{L}_z = \hat{R}_z = L_z = R_z = 0\ .
\end{align}
In the homogeneous ansatz, the action reads
\begin{align}
  S=V_4 \int dz\, \cL=V_4 \int dz\,\left(\cL_g+\cL_{\rm{CS}}+\cL_{\Phi}\right) \ ,
\end{align}
where
\begin{align}
  \cL_g&=-M_5a(z)\left(3H^4(z)+3H^{\prime 2}(z)-\ha_0^{\prime 2}(z)\right) \ ,
\\
\cL_{\rm{CS}}&=\frac{3N_c}{8\pi^2}\,\ha_0(z)H^2(z)H^\prime(z)\ ,
\\
\cL_{\Phi}&=-M_5a^3(z)\,\left(
\frac{3}{2}H^2(z)\omega_0^2(z)+\frac{1}{2}\omega_0^{\prime 2}(z)
+\frac{M_{\Phi}^2}{2}a^2(z)\omega_0^2(z)
+\frac{g}{16}a^2(z)\omega_0^4(z)\right) \ ,
\end{align}
and $V_4=\int d^4 x=V\int dt$.
A prime denotes a derivative with respect to $z$.

The equations of motion (EoMs) are
	\begin{align}
	-2M_5\partial_z\left(a(z)\widehat{a}'_0(z)\right) +\frac{3N_c}{8\pi^2}H^2(z)H'(z)&=0,\label{eqa0}\\
	6M_5 \partial_z\left(a(z)H'(z)\right) -12a(z) M_5H^3(z) -\frac{3N_c}{8\pi^2}\widehat{a}'_0(z)H^2(z) - 3M_5a^3(z)H(z)\omega_0^2(z)&=0,\label{eqH}\\
	M_5\partial_z\left(a^3(z)\omega'_0(z)\right) - 3M_5a^3(z)H^2(z)\omega_0(z)-M_5M_{\Phi}^2a^5(z)\omega_0(z)-\frac{M_5}{4}ga^5(z)\omega_0^3(z)
&=0.\label{eqomega}
	\end{align}
It is found that the solution to the EoM behaves around 
$z=0$ as
\begin{align}
\ha_0(z)&=\mu+\sfa_2\,z^2+\cO(z^4) \ ,~~
\nn\\
  H(z)&=\phi+\sfh_2\,z^2+\phi\left(\frac{m^2}{4}+\phi^2\right)z^2\log z
+\cO(z^4) \ ,
\label{asymUV}
\\
\omega_0(z)&=mz+\sfw_2 z^3+\frac{m}{8}\left(
gm^2+12\phi^2\right)z^3\log z+\cO(z^5) \ .\nn
\end{align}
The GKP-W dictionary
of the AdS/CFT correspondence \cite{Gubser:1998bc,Witten:1998qj} implies 
that the UV boundary values $\mu,\phi$ and $m$ are identified
with the baryon number chemical potential, an axial-isovector
potential, 
and the quark mass, respectively.
The EoM leaves the parameters $\sfa_2,\sfh_2$, and $\sfw_2$ undetermined,
which can be fixed as a function of the boundary
value by imposing an appropriate IR boundary condition at $z=\zir$.

It is useful to note that (\ref{eqa0}) can be integrated as
\begin{align}
  2M_5a(z)\partial_z\ha_0(z)
-\frac{N_c}{8\pi^2}H^3(z)
=
4M_5\sfa_2-\frac{N_c}{8\pi^2}\phi^3 \ .
\label{inteom}
\end{align}
Here, the integration constant is determined by the asymptotic solution
(\ref{asymUV}).

On the basis of the GKP-W dictionary,
we evaluate the on-shell action of the model by inserting the 
solutions to the EoM.
As is well known, the on-shell action is not regular because 
the $z$ integral is divergent at $z=0$. 
In order to regularize it, we set a cut-off
surface at $z=\epsilon>0$.
The finite on-shell action is defined by adding a UV boundary action at
the cut-off surface, which plays a role of the local
counterterm that subtracts the divergence from
the regularized on-shell action. 
This is the procedure for holographic renormalization
and is demonstrated in detail in the next subsection.
For a review, see \cite{hrg}.

In addition to the UV action, we add an IR boundary action at $z=\zir$.
We take 
\begin{align}
  S_{\rm{IR}} =& V_4\,\cL_{\rm{IR}}\big|_{\zir} \ ,~~~\notag
\\
-\cL_{\rm{IR}} =& \,a^4(z)\left(\frac{3}{2}k_1\,a^{-2}(z)H^{2}\omega_0^2
+\frac{k_2}{2}a^{-4}(z)H^4
+\frac{1}{2}m_b^2\omega_0^2+\frac{\lambda}{16}\omega_0^4\right) \ .
\end{align}
Here, $k_1,k_2,m_b$ and $\lambda$ are constants.
This Lagrangian is 
written in terms of only the 5D field $\cL_{\mu},\cR_{\mu}$, and $\Phi$ 
in a manner consistent with $U(2)\times U(2)$ gauge symmetry and 
4D diffeomorphism acting on the IR boundary.
$S_{\rm{IR}}$ specifies the IR boundary conditions obeyed by $H$ and $\omega_0$.

We comment on the physical chemical potential 
and the physical axial-isovector potential.
The UV boundary value $\mu=\ha_0(0)$ 
is not a physical
chemical potential
because it is not gauge invariant
{See also~\cite{
Karch:2009zz,Jensen:2010vd,Hoyos:2021uff} for a related study.}
It is natural to define
the physical chemical potential $\hat{\mu}$ in the gauge invariant
manner as
\begin{align}
  \hat{\mu}=\int_0^{\zir}dz\,\hat{F}_{0z}
=\mu-\ha_0(\zir)\ 
\end{align}
where $\hat{F}_{MN}$ is the field strength of the $U(1)$ vector gauge field.
{Gauge invariant results depend only on the difference $\hat{\mu} = \mu-\ha_0(\zir)$, and the value of $\ha_0(\zir)$ can be chosen arbitrarily due to this property.}
The same reasoning holds for the definition of 
the physical axial-isovector potential $\hat{\phi}$,
which should be defined as
\begin{align}
  \int_0^{\zir}dz\,{ R}_{iz}
=\left(\phi-H(\zir)\right)\frac{\tau_i}{2}
=\hat{\phi}\,\frac{\tau_i}{2}
\end{align}
in the gauge fixing condition $R_z=0\ (L_z=0)$.
The hedgehog structure of the homogeneous ansatz
leaves $\hat{\phi}$ unchanged under the combined transformation
of vectorlike $SU(2)$ and spatial rotation.
{Note that the definition of $\hat{\phi}$ is a conventional one, since it is gauge covariant rather than gauge invariant.}
Throughout this paper, we focus on only the case $\hat{\phi}=0$ 
because this is expected to occur in cases such as the interior of a neutron star.

\subsection{Holographic renormalization and GKP-W relation} 

The regularized on-shell action is obtained by inserting the solution
into the Lagrangian density and integrating it over $\epsilon\le z\le\zir$.
The divergent terms can be extracted only from
the asymptotic solution (\ref{asymUV}) as
\begin{align}
  \ol \cL_g&\equiv\int_{\epsilon}^{\zir}dz\,\cL_{g}\big|_{\rm{EoM}}
=3M_5\phi^4\log\epsilon+\cdots \ ,
\\
  \ol \cL_{\rm{CS}}&\equiv\int_{\epsilon}^{\zir}dz\,\cL_{\rm{CS}}\big|_{\rm{EoM}}
=\mbox{finite} \ ,
\\
  \ol\cL_\Phi&\equiv\int_{\epsilon}^{\zir}dz\,\cL_{\Phi}\big|_{\rm{EoM}}
=M_5\left(
\frac{m^2}{2\epsilon^2}+\frac{3g}{16}m^4\log\epsilon
+3m^2\phi^2\log\epsilon\right)+\cdots \ .
\end{align}
The total regularized Lagrangian is 
\begin{align}
  \cL_{\rm{reg}}\equiv \ol\cL_g+\ol\cL_{\rm{CS}}+\ol\cL_{\Phi}+\ol\cL_{\rm{IR}}
\ ,
\end{align}
with $\ol\cL_{\rm{IR}}\equiv\cL_{\rm{IR}}|_{\rm{EoM}}$ finite.
The UV divergence can be canceled by adding the counterterm
\begin{align}
  \cL_c=-M_5\left(
\frac{m^2}{2\epsilon^2}+\frac{3g}{16}m^4\log\epsilon
+3m^2\phi^2\log\epsilon
+3\phi^4\log\epsilon
\right) \ .
\end{align}
Here, we adopt the minimal subtraction scheme, where 
$\cL_c$ contains no $\cO(\epsilon^0)$ term.
The subtracted Lagrangian is  defined by 
\begin{align}
  \cL_{\rm{sub}}=\cL_{\rm{reg}}+\cL_c \ ,
\end{align}
from which the renormalized on-shell action is obtained by 
${\displaystyle V_4\cdot\lim_{\epsilon\to 0}\cL_{\rm{sub}}}$.

In a recipe for holographic renormalization, we write
the UV boundary values $\phi,\mu$, and $m$ as a function of
$\ha_0(\epsilon),H(\epsilon)$, and $\omega_0(\epsilon)$.
It is shown that
\begin{align}
\mu&=\ha_0(\epsilon)+\cO(\epsilon^2) \ ,
\\
  \phi&=H(\epsilon)+\cO(\epsilon^2)\ ,
\\
m&=\frac{1}{\epsilon}\omega_0(\epsilon)
-\epsilon^2\,\sfw_2-\frac{\log\epsilon}{8\epsilon}\,g\,\omega_0^3(\epsilon)
-\frac{3}{2}\epsilon\log\epsilon\,\omega_0(\epsilon)H^2(\epsilon)
+\cO(\epsilon^4) \ .
\end{align}
Then, the counterterm becomes
\begin{align}
  \cL_c=-M_5\left(
\frac{1}{2\epsilon^4}\,\omega_0{{^2}}(\epsilon)
+\frac{3\log\epsilon}{2\epsilon^2}\,\omega^2_0(\epsilon)\,H^2(\epsilon)
+\frac{\log\epsilon}{16\epsilon^4}\,g\,\omega_0^4(\epsilon)
+3\log\epsilon\,H^4(\epsilon)
\right)
+\cO(\epsilon^0)\ .
\end{align}
{ Note that this is consistent with $U(2)\times U(2)$ gauge 
symmetry and 4D diffeomorphism on the cut-off surface.}

It follows from the Hamilton-Jacobi equation that
the variation of the subtracted on-shell Lagrangian 
with respect to that of the
boundary values reads 
\begin{align}
  \delta\cL_{\rm{sub}}=
  -\left(p_{\ha_0}(\epsilon)-\frac{\partial\cL_c}{\partial \ha_0(\epsilon)}\right)
\delta \ha_0(\epsilon)\ 
-\left(p_H(\epsilon)-\frac{\partial\cL_c}{\partial H(\epsilon)}\right)
\delta H(\epsilon)
-\left(p_{\omega_0}(\epsilon)-\frac{\partial\cL_c}{\partial \omega_0(\epsilon)}\right)
\delta \omega_0(\epsilon).
\end{align}
Here,
\begin{align}
  p_{\ha_0}&=\frac{\partial\cL}{\partial \ha_0^{\prime}}
=
2M_5a\,\ha_0^\prime \ ,
\nn\\
  p_H&=\frac{\partial\cL}{\partial H^{\prime}}
=
-6M_5aH^\prime+\frac{3N_c}{8\pi^2}\ha_0H^2 \ ,
\nn\\
  p_{\omega_0}&=\frac{\partial\cL}{\partial \omega_0^{\prime}}
=
-M_5a^3\omega_0^\prime \ .\nn
\end{align}
Note that the variation comes only from the UV boundary
irrespective of the IR boundary conditions imposed.
See Appendix \ref{HJeqn} for details.
It is found that
\begin{align}
  p_{\ha_0}(\epsilon)&=4M_5\sfa_2+\cO(\epsilon^2) \ ,\nn\\
  p_H(\epsilon)&=
-12M_5\sfh_2-6M_5\phi\left(\frac{m^2}{4}+\phi^2\right)
(2\log\epsilon+1)
+\frac{3N_c}{8\pi^2}\mu\phi^2+\cO(\epsilon) \ ,
\nn\\
  p_{\omega_0}(\epsilon)&=
-\frac{mM_5}{\epsilon^3}-\frac{3M_5\sfw_2}{\epsilon}
-\frac{mM_5}{8\epsilon}\left(3\log\epsilon+1\right)
\left(gm^2+12\phi^2\right)
+\cO(\epsilon) \ ,\nn
\end{align}
and
\begin{align}
  \frac{\partial\cL_c}{\partial \ha_0(\epsilon)}&=0 \ ,\nn\\
  \frac{\partial\cL_c}{\partial H(\epsilon)}&=
-3M_5\phi(m^2+4\phi^2)\log\epsilon \ ,
\nn\\
  \epsilon\frac{\partial\cL_c}{\partial \omega_0(\epsilon)}&=
-\frac{mM_5}{\epsilon^2}
-\frac{3mM_5}{8}\log\epsilon\,
(gm^2+12\phi^2)
\ .\nn
\end{align}
We obtain
\begin{align}
  d&\equiv
\lim_{\epsilon\to 0}\frac{\partial \cL_{\rm{sub}}}{\partial \ha_0(\epsilon)}
=
-\lim_{\epsilon\to 0}\left(p_{\ha_0}(\epsilon)-\frac{\partial\cL_c}{\partial \ha_0(\epsilon)}\right)=-4M_5\sfa_2 \ ,
\label{gkpw:d}\\
  J&\equiv
\lim_{\epsilon\to 0}\frac{\partial \cL_{\rm{sub}}}{\partial H(\epsilon)}
=
-\lim_{\epsilon\to 0}\left(p_H(\epsilon)-\frac{\partial\cL_c}{\partial H(\epsilon)}\right)
=12M_5\sfh_2+6M_5\phi
\left(\frac{m^2}{4}+\phi^2\right)-\frac{3N_c}{8\pi^2}\mu\phi^2 \ ,
\label{gkpw:J}
\\
  \xi&\equiv
\lim_{\epsilon\to 0}\epsilon\frac{\partial \cL_{\rm{sub}}}{\partial \omega_0(\epsilon)}
=
-\lim_{\epsilon\to 0}\epsilon\left(p_{\omega_0}(\epsilon)-\frac{\partial\cL_c}{\partial \omega_0(\epsilon)}\right)
=3M_5\sfw_2+\frac{mM_5}{8}(gm^2+12\phi^2) \ .
\label{gkpw:xi}
\end{align}
These are the GKP-W relations for a holographic model of QCD
in the homogeneous ansatz.

Normalizing the baryon number density so that a single baryon
carries a unit baryon number, we find
\begin{align}
  d_B&=\frac{2}{N_c}d
=-\frac{8M_5}{N_c}\sfa_2
\nn\\
&=
\frac{1}{4\pi^2}(H^3(\zir)-\phi^3)
-\frac{4M_5}{N_c}\,a(\zir)\,\ha_0^{\prime}(\zir)\ .
\label{dB}
\end{align}
Here, (\ref{inteom}) is used to eliminate $\sfa_2$. 
The corresponding baryon number chemical potential $\mu_B$ is
given by noting
$\hat{\mu} d=\mu_B d_B$. We obtain
\begin{align}
\mu_B=\frac{N_c}{2}\hat{\mu} \ .  
\end{align}

Some comments about  $d_B$  are in order.
As shown below, the baryonic matter phase is realized by imposing
the Dirichlet boundary condition for $\hat{a}_0$ on the IR boundary.
This implies that $\ha_0'(\zir)\ne 0$. 
Furthermore, we turn off the physical axial-isovector potential
$\hat{\phi}=0$. Then, it follows from (\ref{dB}) that
only $\ha_0'(\zir)$ contributes to the baryon number density.
A more direct way for computing $d_B$ is to evaluate $\sfa_2$ without
relying on (\ref{dB}).

\subsection{IR boundary condition}

The IR boundary condition is determined by requiring that the variation of the action $S+S_{\rm{IR}}$ vanish at $z=\zir$. We first discuss the IR Neumann boundary condition.
Using
\begin{align}
 \frac{\partial\cL_{\rm{IR}}}{\partial \ha_0}
&=0 \ ,
\nn\\
 \frac{\partial\cL_{\rm{IR}}}{\partial H}
&=
-3k_1a^2\,H\omega_0^2
-2k_2H^3
\ ,
\nn\\
 \frac{\partial\cL_{\rm{IR}}}{\partial \omega_0}
&=
-a^4\left[3{ k_1}a^{-2}\,H^2\omega_0+m_b^2\omega_0+\frac{\lambda}{4}\omega_0^3
\right]\ ,\nn
\end{align}
the IR Neumann boundary condition for $\ha_0, H$, and $\omega_0$ 
reads
\begin{align}
0&=2M_5a\,\ha_0^\prime \ ,
\label{IRN:ha0}
\\
  0&=
-6M_5aH^\prime+\frac{3N_c}{8\pi^2}\ha_0H^2
{
-3k_1a^2\,H\omega_0^2
-2k_2H^3} 
 \ ,
\label{IRN:H}
\\
0&=-M_5a^3\omega_0^\prime
-a^4\left[3{ k_1}a^{-2}\,H^2\omega_0+m_b^2\omega_0+\frac{\lambda}{4}\omega_0^3
\right] \ ,
\label{IRN:w}
\end{align}
respectively.
In terms of $\omega_0(z)=\bar{\omega}_0(z)/a(z)$, (\ref{IRN:w}) is rewritten as
\begin{align}
  0=M_5 a\left(\bar{\omega}_0^\prime+a\,\bar{\omega}_0
\right)
+
3{ k_1}H^2\bar{\omega}_0+a^2\left(m_b^2\bar{\omega}_0+\frac{\lambda}{4}{a^{-2}}\bar{\omega}_0^3\right)
\ .
\end{align}
It turns out to be more useful to work with $\ol\omega_0$ rather
than $\omega_0$ 
when solving the EoM numerically.

The Dirichlet boundary condition for $\ha_0,H$ and $\bar{\omega}_0$ is
given by
\begin{align}
  \ha_0(\zir)=A \ ,~~~
  H(\zir)=B \ ,~~~
  \ol\omega_0(\zir)=C \ .
\label{IRD}
\end{align}
Here, $A,B$, and $C$ are the IR boundary values that are chosen by hand so that
the solution is consistent with the experimental results.

The solution to the EoM is obtained by suitably choosing the IR boundary
conditions. We argue that each solution defines a distinct QCD phase,
and the phase structure as a function of the baryon
number chemical potential is determined energetically by comparing 
the grand potential of the phases.
The grand potential of a phase is identified with the on-shell action
that is evaluated by inserting the corresponding solution of EoM
into the action.
This rule is in parallel with the holographic description of the 
confinement/deconfinement phase transition in terms of the
Hawking-Page transition \cite{Witten:1998zw}.
Naively, there exist eight QCD phases because
eight IR boundary conditions are allowed to impose. 
We propose, however,
that $\ha_0$ and $H$ must have a common IR boundary condition, 
that is, either the Neumann or the Dirichlet boundary condition.
This is understood naturally from the 4D Lorentz symmetry by noting 
that
$\ha_0$ and $H$ come from the temporal and
the spatial components of the 5D gauge field, respectively, 
see (\ref{homo:ansatz}).

First, we solve the EoM of $H$ by imposing the Neumann boundary 
condition (\ref{IRN:w}) at $z=\zir$.
The solution is given simply by
\begin{align}
  H(z)=0 \ .
\end{align}
Then, the general solution of $\ha_0$ and $\omega_0$ is given by
\begin{align}
  \ha_0(z)=\mu+\sfa_2 z^2 \ ,~~~
\omega_0(z)=mz+\sfw_2 z^3 \ .
\end{align}
Note that this is consistent with the UV asymptotic behavior (\ref{asymUV}).
In this case, we have two boundary conditions to impose for
$\hat{a}_0$ and $\omega_0$:
\begin{enumerate}[label=(\roman*)]
  \item NNN-type, which states that $\ha_0$, $H$, and $\omega_0$ all
obey the IR Neumann boundary condition. It follows from (\ref{IRN:ha0}) that
\begin{align}
  \sfa_2=0 \ ,
\end{align}
showing that the baryon number density vanishes because of (\ref{gkpw:d}).
Using the IR boundary condition (\ref{IRN:w}) gives 
\begin{align}
  \frac{\lambda}{4}\left(\wt \sfw_2+\wt m\right)^3
+(m_b^2+3M_5)\left(\wt \sfw_2+\wt m\right)-2M_5\wt m=0 \ .
\label{peqn:w2}
\end{align}
Here,
\begin{align}
\wt m=m\zir \ ,~~~  \wt\sfw_2=\sfw_2\zir^3 \ .\nn
\end{align}
This allows a nonvanishing $\sfw_2$, leading to
a nonvanishing condensate $\xi$ because of (\ref{gkpw:xi}).
The NNN-type solution is a candidate for the
QCD vacuum for $\mu_B=0$. 
If this is the case,
$m_b^2$ is fixed by solving (\ref{peqn:w2})
as a function of the experimental input for the 
quark mass $m$ and the chiral condensate $\xi_0$,
\begin{align}
  \left(m_b^{\rm{NNN}}\right)^2=-3M_5
+\frac{2 M_5\wt m}{\wt m+\wt\sfw_2^{(0)}}
-\frac{\lambda}{4}\left(\wt m+\wt\sfw_2^{(0)}\right)^2 \ ,
\label{mb2nnn}
\end{align}
where $\sfw_2^{(0)}$ is written in terms of
$\xi_{0}$ using the GKP-W relation (\ref{gkpw:xi}).

Since (\ref{peqn:w2}) is a cubic polynomial in $\sfw_2$, there
might be two more solutions that are in conflict with
the experimental data. In order to examine them, 
we note that (\ref{peqn:w2}) is rewritten as
\begin{align}
-\frac{\lambda}{4}\left(\wt m+\wt\sfw_2\right)^2 
+\frac{2 M_5\wt m}{\wt m+\wt\sfw_2}
=
-\frac{\lambda}{4}\left(\wt m+\wt\sfw_2^{(0)}\right)^2 
+\frac{2 M_5\wt m}{\wt m+\wt\sfw_2^{(0)}}\ .\nn
\end{align}
The solution to this equation other than $\wt\sfw_2=\wt\sfw_2^{(0)}$ is 
given by
\begin{align}
  \wt m+\wt\sfw_2
=
\frac{\wt m+\wt\sfw_2^{(0)}}{2}
\left[
-1\pm\sqrt{1-\frac{32M_5\wt m}{\lambda(\wt m+\wt\sfw_2^{(0)})^3}}
\,\right]\ .\nn
\end{align}
We require 
\begin{align}
  1-\frac{32M_5\wt m}{\lambda(\wt m+\wt\sfw_2^{(0)})^3}<0 \ .
\label{ineq:lam}
\end{align}
This guarantees that the IR Neumann boundary condition (\ref{IRN:w})
singles out only the physical value of the chiral condensate.

\item
NND-type, which states that $\ha_0$ and $H$ obey the Neumann boundary
condition at $z=\zir$ and $\omega_0$ Dirichlet boundary condition given
in (\ref{IRD}). It follows that
\begin{align}
  \sfa_2=0 \ ,~~~
\wt m+\wt\sfw_2=\wt C \ ,\nn
\end{align}
with $\wt C=C\zir$.
This is another candidate for the QCD vacuum for $\mu_B=0$.
As in the NNN-type solution, we give the input for the quark mass
and the chiral condensate. Then, the IR boundary value $C$ is fixed
as
\begin{align}
  \wt C^{\rm{NND}}=\wt m+\wt\sfw_2^{(0)}\ .\nn
\end{align}
In this case, the parameter $m_b^2$ appearing in the IR boundary action
remains unfixed.

\end{enumerate}

The GKP-W dictionary shows that
the grand potential $\Omega$ of a classical solution 
is given by the on-shell action
\begin{align}
\frac{\Omega}{V}=-\lim_{\epsilon\to 0}\left(\ol\cL_g+\ol\cL_{\rm{CS}}
+\ol\cL_\Phi+\ol\cL_{\rm{IR}}+\cL_c\right) \ .
\label{gkp:omega}
\end{align}
For both cases, the on-shell Lagrangians
$\ol\cL_g$ and $\ol\cL_{\rm{CS}}$ vanish, and
\begin{align}
 \lim_{\epsilon\to 0}\left( \ol\cL_{\Phi}+\cL_c\right)
=
-\frac{M_5}{2\zir^4}\left(
\wt m^2+3\left(\wt\sfw_2^{(0)}\right)^2
\right) \ .\nn
\end{align}
For the NNN case, the on-shell IR boundary action reads
\begin{align}
  \ol\cL_{\rm{IR}}^{\rm{NNN}}=
-\frac{1}{2\zir^4}\left[
\left(m_b^{\rm{NNN}}\right)^2\left(\wt m+\wt\sfw_2^{(0)}\right)^2
+\frac{\lambda}{8}\left(\wt m+\wt\sfw_2^{(0)}\right)^4\right]\ .
\end{align}
For the NND case, we have
\begin{align}
  \ol\cL_{\rm{IR}}^{\rm{NND}}=
-\frac{1}{2\zir^4}\left[
\left(m_b^{\rm{NND}}\right)^2\left(\wt m+\wt\sfw_2^{(0)}\right)^2
+\frac{\lambda}{8}\left(\wt m+\wt\sfw_2^{(0)}\right)^4\right]\ ,
\end{align}
with $m_b^{\rm{NND}}$ left unknown.
It then follows that the grand potential  is given by the same expression for both NNN and NND cases,
\begin{align}
\frac{\Omega^{\rm{NNN\, or\, NND}}}{V} =
\frac{M_5}{2\zir^4}\left(
\wt m^2+3\left(\wt\sfw_2^{(0)}\right)^2
\right)
+
\frac{1}{2\zir^4}\left[
\left(m_b^{\rm{NNN\, or\, NND}}\right)^2\left(\wt m+\wt\sfw_2^{(0)}\right)^2
+\frac{\lambda}{8}\left(\wt m+\wt\sfw_2^{(0)}\right)^4\right]
\ ,
\end{align}
where only the mass parameter $m_b$ differs for NNN and NND.
Note that this is independent of the chemical potential $\mu_B$.

Next, we solve $H(z)$ by imposing the Dirichlet boundary condition
at $z=\zir$
\begin{align}
  H(\zir)=B \ .\nn
\end{align}
Then, there remain two more IR boundary conditions to impose:
\begin{enumerate}[resume*]
  \item 
DDN-type, which states that $\ha_0$ and $H$ obey the
Dirichlet boundary condition and $\omega_0$ the Neumann boundary condition,
\begin{align}
  \ha_0(\zir)=A \ ,~~H(\zir)=B \ ,~~
M_5a^3\omega_0^\prime
+a^4\left[3{ k_1}a^{-2}\,H^2\omega_0+m_b^2\omega_0+\frac{\lambda}{4}\omega_0^3
\right]=0 \ .
\end{align}

\item
DDD-type, which states that $\ha_0,H$ and $\omega_0$ all
obey the Dirichlet boundary condition (\ref{IRD}).
\end{enumerate}

For {all the above cases}, we solve the EoMs (\ref{eqa0}), (\ref{eqH}), and
(\ref{eqomega}) numerically by using the shooting method.
For a given $\mu,\phi$, and $m$, we give a test value to $\sfa_2,\sfh_2$,
and $\sfw_2$. This enables us to solve the EoMs numerically by imposing
the initial condition near $z=0$, which is derived from
the UV asymptotic behavior (\ref{asymUV}).
Then, we examine if the resultant numerical solution obeys
the IR boundary condition at $z=\zir$.
If not, we give a new test value to 
$\sfa_2,\sfh_2$, and $\sfw_2$ and repeat the procedures outlined above
until we reach the correct result of $\sfa_2,\sfh_2$, and $\sfw_2$.

\section{Numerical calculations and results}

We first discuss how the parameters in the holographic model are
fixed.
The quartic potential term of the complex scalar $\Phi$ is turned off 
by setting $g=0$ for simplicity.
We fix $m$ to be
the up and down quark masses 
$m=3 \mathrm{\,MeV}$.
In addition, the number of colors is $N_c=3$.
The IR boundary action contains 
four parameters $k_1,k_2,m_b$ and $\lambda$.
We set {$k_1=0$} for simplicity. 
Note that we examined the effects of varying $k_1$ in several cases of interest and found that they are not significant.
In order to fix $m_b$ and $\lambda$, we propose that 
the QCD vacuum for $\mu_B=0$ is described by
the NNN-type solution. This is equivalent to 
imposing 
\begin{align}
  m_b^{\rm{NND}}>m_b^{\rm{NNN}} \ ,
\end{align}
which guarantees that
the NNN-type solution is energetically favored over the NND-type solution.
As an experimental input, we set the chiral condensate
$\xi_0=(251\ \mathrm{MeV})^3$, which is obtained 
from a lattice calculation \cite{JLQCD:2007ppn}.
Using this and (\ref{mb2nnn}) fixes
the boundary mass $m_b$.
It is seen that, for the 
parameters chosen above, a natural solution to 
the inequality (\ref{ineq:lam}) is given by $\lambda=0$.
The IR boundary is located at $z=\zir$, on which the 
IR boundary action is defined. We set $\zir=L$, as in 
\cite{Bartolini:2022rkl}.

As the chemical potential increases, 
the NNN-type solution is expected to persist as the energetically
most favored phase up to a critical
value of $\mu_B$.
We call this phase the {nonbaryonic phase}.
At the critical point, 
the phase transition to a baryonic matter phase
with a nonvanishing baryon number density occurs.
We now examine the DDN-type and DDD-type solutions 
to see if these capture
the expected nature of the baryonic matter phase.

We start analyzing the DDN-type boundary condition.
This is specified with the IR boundary values 
$\hat{a}_0(\zir)=A$ and $H(\zir)=B$.
In this paper, we set $A=4$ in units of $L^{-1}$. 
{This is because the choice of the value of $A$ does not affect the results, owing to the gauge invariance in the bulk. It is also confirmed that the numerical results are not sensitive to the specific choice of $A$.}
We focus on the case $\hat{\phi}=0$ for simplicity so that $\phi=B$.
Then, there remains the {four free parameters $\mu_B,B,L$, and $k_2$.}
Because there is no experimental input 
to determine {$B,L$, and $k_2$,} we make a sample 
computation for some values of them chosen by hand.
In this paper, we take
(i) $L^{-1}=323\ \mathrm{MeV}$,~$B=0.4/L,\ 0.6/L,\ 0.8/L$, and
(ii) $L^{-1}=170\ \mathrm{MeV}$,~$B=0.4/L,\ 0.6/L,\ 0.8/L$.
Setting $L^{-1}=323\ \mathrm{MeV}$ is argued to achieve the best fit of the predictions for
the meson sector with experiments \cite{Erlich:2005qh}.
It is seen below that the case (ii) reproduces the critical value of $d_B=n_0$ at $\mu_B=1\ \mathrm{GeV}$ consistently with experiments, where $n_0=0.17\ \mathrm{fm}^{-3}$ is the nuclear saturation density.

\begin{figure}[htbp]
    \begin{tabular}{ccc}
      \begin{minipage}[h]{0.33\columnwidth}
        \centering
        \includegraphics[width=0.95\columnwidth]{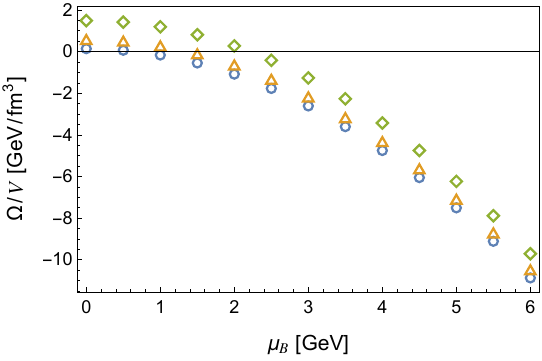}
      \end{minipage} 
      \begin{minipage}[h]{0.33\columnwidth}
        \centering
        \includegraphics[width=0.95\columnwidth]{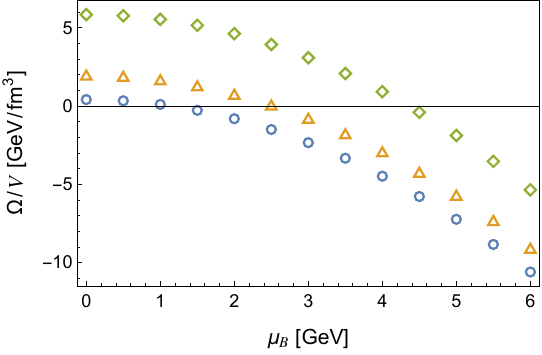}
      \end{minipage} 
      \begin{minipage}[h]{0.33\columnwidth}
        \centering
        \includegraphics[width=0.95\columnwidth]{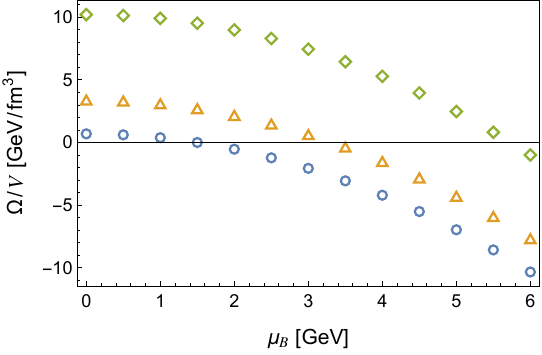}
      \end{minipage}    
      \\\\
     \begin{minipage}[h]{0.33\columnwidth}
        \centering
        \includegraphics[width=0.95\columnwidth]{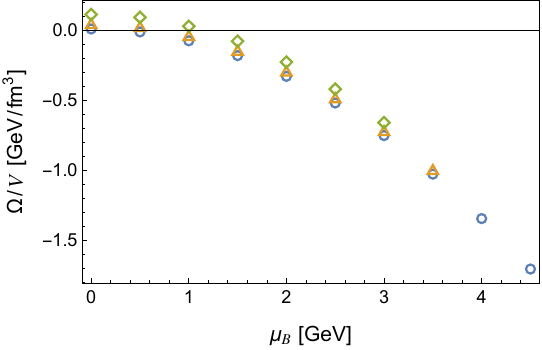}
      \end{minipage} 
      \begin{minipage}[h]{0.33\columnwidth}
        \centering
        \includegraphics[width=0.95\columnwidth]{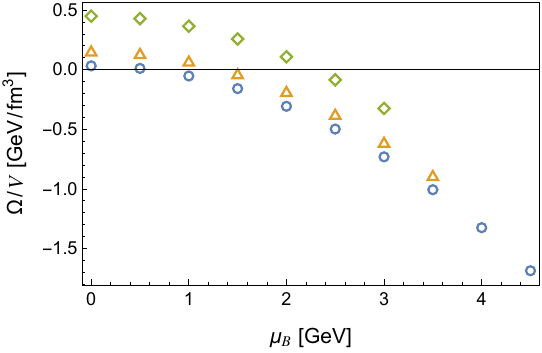}
      \end{minipage} 
      \begin{minipage}[h]{0.33\columnwidth}
        \centering
        \includegraphics[width=0.95\columnwidth]{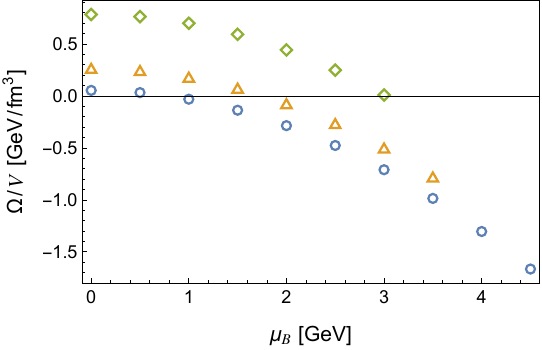}
      \end{minipage}    
      \end{tabular}
       \caption{The grand potential density as a function of $\mu_B$ at $L^{-1}=323\ \mathrm{MeV}$ (upper) and $L^{-1}=170\ \mathrm{MeV}$ (lower) for $k_2=5$ (left), $k_2=20$ (middle), and $k_2=35$ (right).
       Blue circles, orange triangles, and green diamonds are for $B = 0.4/L$, $0.6/L$, and $0.8/L$, respectively.}
        \label{fig:gp}
  \end{figure}

\begin{figure}[htbp]
    \begin{tabular}{cc}
      \begin{minipage}[h]{0.33\columnwidth}
        \centering
        \includegraphics[width=0.9\columnwidth]{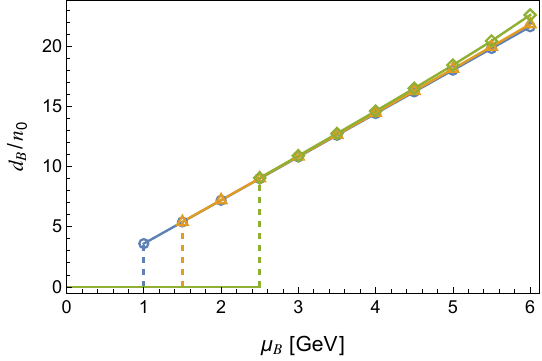}
      \end{minipage} 
      \begin{minipage}[h]{0.33\columnwidth}
        \centering
        \includegraphics[width=0.9\columnwidth]{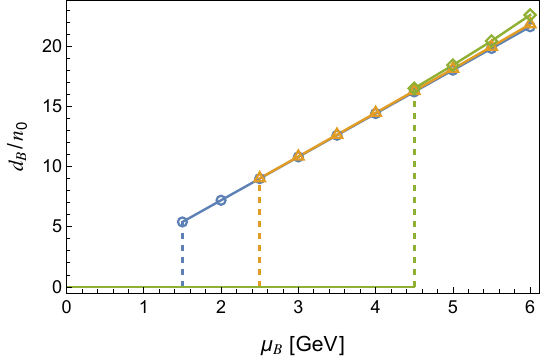}
      \end{minipage} 
      \begin{minipage}[h]{0.33\columnwidth}
        \centering
        \includegraphics[width=0.9\columnwidth]{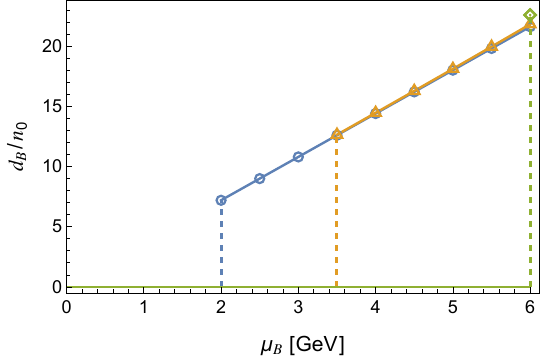}
      \end{minipage}    
      \\\\
      \begin{minipage}[h]{0.33\columnwidth}
        \centering
        \includegraphics[width=0.9\columnwidth]{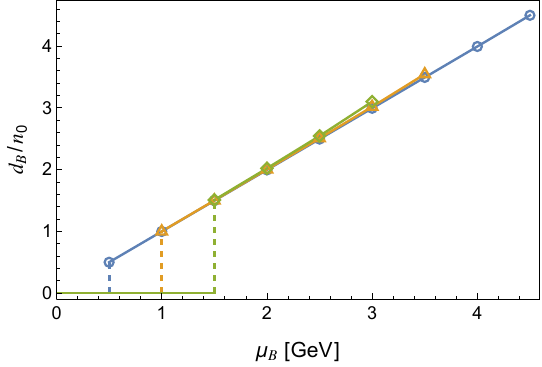}
      \end{minipage} 
      \begin{minipage}[h]{0.33\columnwidth}
        \centering
        \includegraphics[width=0.9\columnwidth]{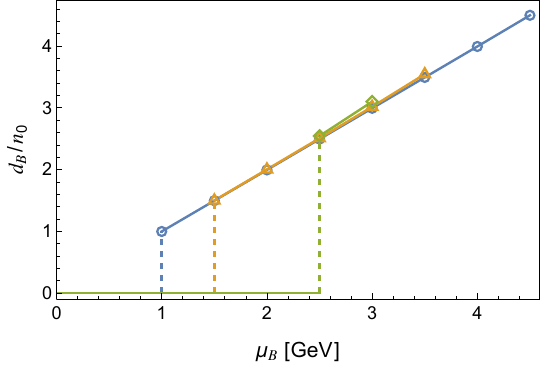}
      \end{minipage} 
      \begin{minipage}[h]{0.33\columnwidth}
        \centering
        \includegraphics[width=0.9\columnwidth]{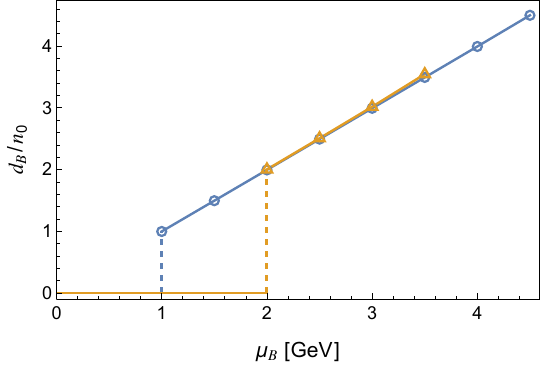}
      \end{minipage}    
      \end{tabular}
       \caption{The baryon number density around $\mu_B=1\ \mathrm{GeV}$ as a function of $\mu_B$ at $L^{-1}=323\ \mathrm{MeV}$ (upper) and $L^{-1}=170\ \mathrm{MeV}$ (lower) for $k_2=5$ (left), $k_2=20$ (middle), and $k_2=35$ (right).
	Blue circles, orange triangles and green diamonds are for $B = 0.4/L$, $0.6/L$ and $0.8/L$, respectively.
	The dashed lines represent the points closest to the critical point among the numerically calculated points.}
        \label{fig:density}
  \end{figure}

\begin{figure}[htbp]
    \begin{tabular}{cc}
      \begin{minipage}[h]{0.33\columnwidth}
        \centering
        \includegraphics[width=0.9\columnwidth]{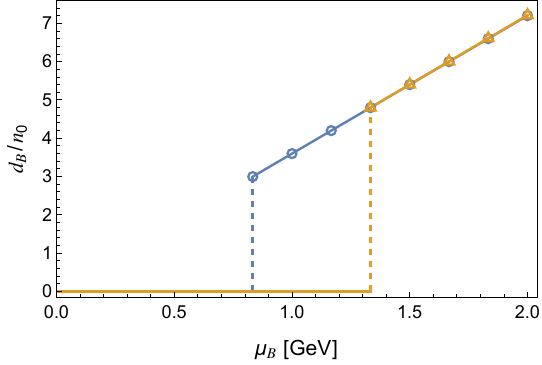}
      \end{minipage} 
      \begin{minipage}[h]{0.33\columnwidth}
        \centering
        \includegraphics[width=0.9\columnwidth]{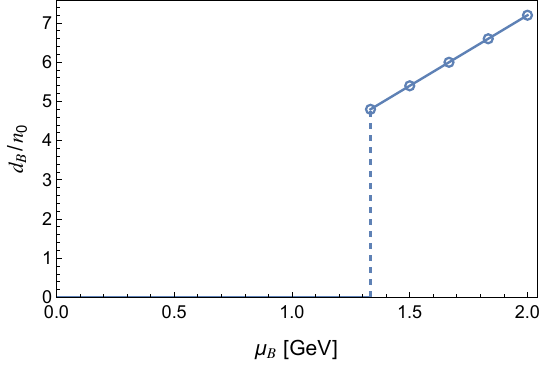}
      \end{minipage} 
      \begin{minipage}[h]{0.33\columnwidth}
        \centering
        \includegraphics[width=0.9\columnwidth]{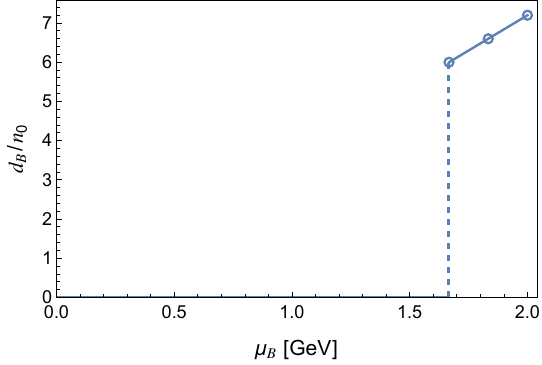}
      \end{minipage}    
      \\\\
      \begin{minipage}[h]{0.33\columnwidth}
        \centering
        \includegraphics[width=0.9\columnwidth]{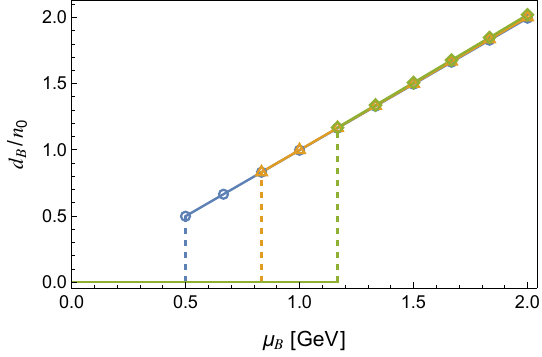}
      \end{minipage} 
      \begin{minipage}[h]{0.33\columnwidth}
        \centering
        \includegraphics[width=0.9\columnwidth]{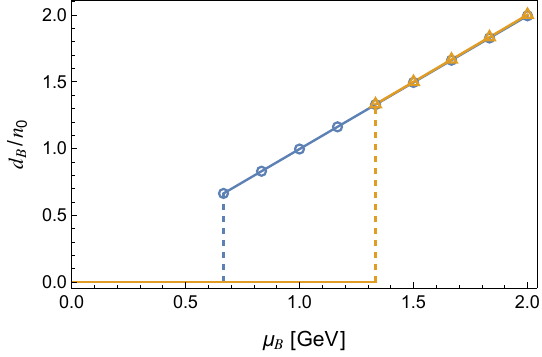}
      \end{minipage} 
      \begin{minipage}[h]{0.33\columnwidth}
        \centering
        \includegraphics[width=0.9\columnwidth]{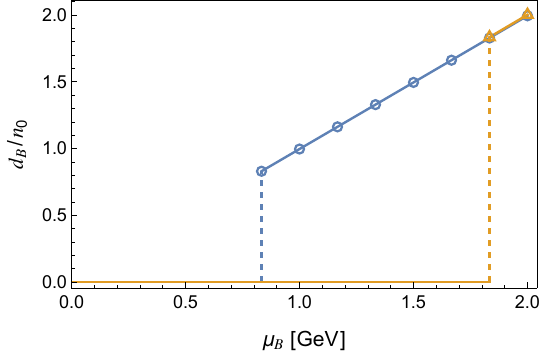}
      \end{minipage}    
      \end{tabular}
       \caption{The baryon number density as a function of $\mu_B$ at $L^{-1}=323\ \mathrm{MeV}$ (upper) and $L^{-1}=170\ \mathrm{MeV}$ (lower) for $k_2=5$ (left), $k_2=20$ (middle), and $k_2=35$ (right).
       Blue circles, orange triangles and green diamonds are for $B = 0.4/L$, $0.6/L$ and $0.8/L$, respectively.
       The dashed lines represent the points closest to the critical point among the numerically calculated points.}
        \label{fig:density2}
  \end{figure}

Now let us compute the grand potential as a function of $\mu_B$.
We solve the EoM numerically by using the shooting method with
the DDN boundary condition imposed at $z=\zir$ to obtain the
grand potential from the on-shell action, see (\ref{gkp:omega}). 
It is useful to define the grand potential by subtracting 
$\Omega^{\rm{NNN}}$ from it.
The results are shown in Fig.~\ref{fig:gp}.
%
These show that the DDN-type solution is more favored energetically
than the nonbaryonic phase for $\mu_B$ larger than the critical value.
We call the phase described by this solution the baryonic phase.
The reason for ``baryonic'' is that
this phase has a nonvanishing baryon number density,
see Fig.~\ref{fig:density}.
Figure~\ref{fig:density2} shows more detailed figures around $\mu_B=1\ \mathrm{GeV}$.

%

As found in Fig.~\ref{fig:gp},
the orange triangles and the green diamonds 
suddenly disappear as the chemical potential increases.
This is a manifestation of the fact that no classical solution
satisfying the DDN boundary condition exists in the corresponding
region.
In particular, for the case of $L^{-1}=170\ \mathrm{MeV}$, $k_2=35$, 
and $B=0.8/L$, the DDN-type solution ceases to exist 
before $\mu_B$ reaches the critical value.
These facts imply that the effective description of the high-density
QCD phases in terms of the homogeneous ansatz is invalid.

For $L^{-1}=323\ \mathrm{MeV}$, the estimated critical values
of $\mu_B$ and $d_B$ tend to be larger than values expected from the nucleon mass in the vacuum, $d_B/n_0\sim 1$ at $\mu_B\sim 1\ \mathrm{GeV}$.
Naively, we observe that the models with 
$L^{-1}=323\ \mathrm{MeV}$ are unnatural from the viewpoint of the
hadron physics if we assume that the nonbaryonic phase undergoes
a phase transition to the baryonic phase directly.
In the next section, we propose an alternative scenario of the 
phase transition where the DDN-type solution with 
$L^{-1}=323\ \mathrm{MeV}$ plays a role in describing a  higher-density
phase.

For $L^{-1}=170\ \mathrm{MeV}$, 
the baryon number density increases as a function of $\mu_B$
much more slowly than what is expected from the empirical knowledge of nuclear physics.
See~\cite{Brandes:2023hma}, for example.

\begin{figure}[htbp]
    \begin{tabular}{cc}
      \begin{minipage}[h]{0.33\columnwidth}
        \centering
        \includegraphics[width=0.9\columnwidth]{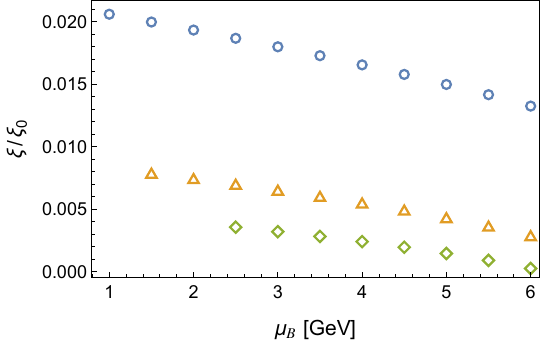}
      \end{minipage} 
      \begin{minipage}[h]{0.33\columnwidth}
        \centering
        \includegraphics[width=0.9\columnwidth]{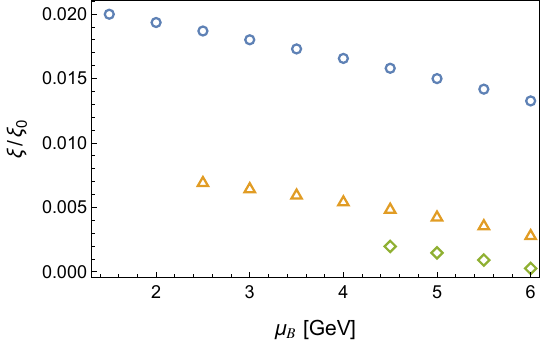}
      \end{minipage} 
      \begin{minipage}[h]{0.33\columnwidth}
        \centering
        \includegraphics[width=0.9\columnwidth]{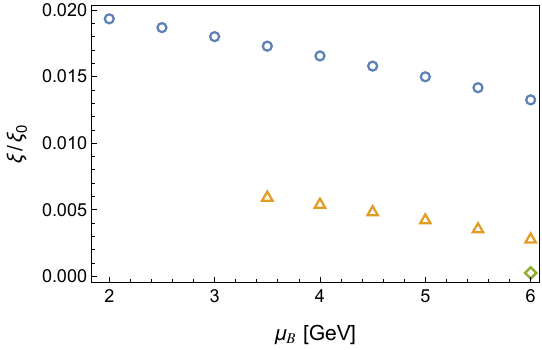}
      \end{minipage}    
      \\\\
      \begin{minipage}[h]{0.33\columnwidth}
        \centering
        \includegraphics[width=0.9\columnwidth]{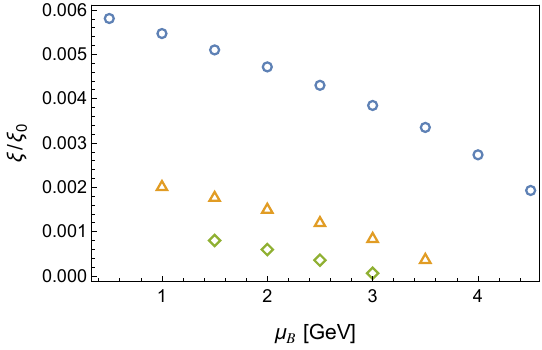}
      \end{minipage} 
      \begin{minipage}[h]{0.33\columnwidth}
        \centering
        \includegraphics[width=0.9\columnwidth]{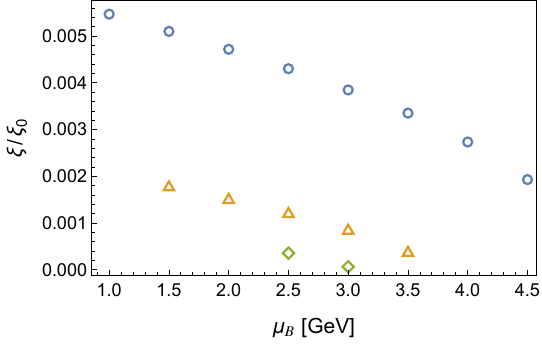}
      \end{minipage} 
      \begin{minipage}[h]{0.33\columnwidth}
        \centering
        \includegraphics[width=0.9\columnwidth]{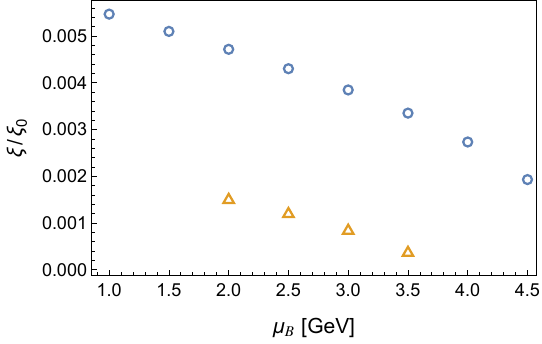}
      \end{minipage}    
      \end{tabular}
       \caption{The chiral condensate normalized by the value in the nonbaryonic phase $\xi_0$ as a function of $\mu_B$ at $L^{-1}=323\ \mathrm{MeV}$ (upper) and $L^{-1}=170\ \mathrm{MeV}$ (lower) for $k_2=5$ (left), $k_2=20$ (middle), and $k_2=35$ (right).
       Blue circles, orange triangles and green diamonds are for $B = 0.4/L$, $0.6/L$ and $0.8/L$, respectively.}
        \label{fig:xi}
  \end{figure}

Figure~\ref{fig:xi} shows the chiral condensate in the baryonic phase $\xi$ 
normalized by $\xi_0$ as a function of $\mu_B$.
It is found that the chiral condensate is much smaller than $\xi_0$ and decreases as $\mu_B$ grows in the baryonic phase.
The solutions of the equation of motion satisfying the DDN boundary condition are not allowed to exist after the chiral condensate crosses zero.

We make some comments about
the DDD-type IR boundary condition.
In addition to $A$ and $B$, this boundary condition is characterized by
another boundary value $C$.  
It is found numerically that the DDD-type solution describes a phase
very similar to the DDN-type solution---it carries a nonvanishing
baryon number density and exhibits partial restoration of
chiral symmetry. 
Hence, the DDD-type solution provides us with another candidate for the baryonic matter phase in the holographic setup.
In this paper, we assume that the DDN-type solution is energetically more favored than the DDD-type solution.
This is achieved by choosing an appropriate value of $C$ so that the grand potential density for the DDN-type solution is lower than that of the DDD-type solution.

\subsection{Equation of state, speed of sound and axial-isovector condensate}

\begin{figure}[htbp]
    \begin{tabular}{cc}
      \begin{minipage}[h]{0.33\columnwidth}
        \centering
        \includegraphics[width=0.9\columnwidth]{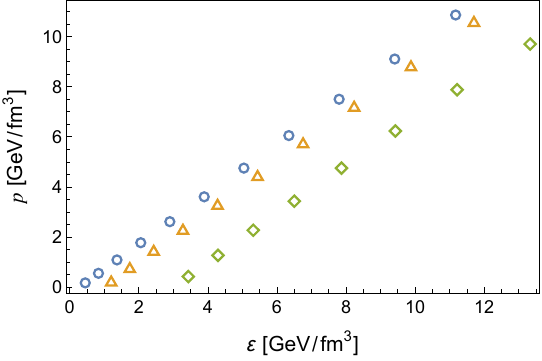}
      \end{minipage} 
      \begin{minipage}[h]{0.33\columnwidth}
        \centering
        \includegraphics[width=0.9\columnwidth]{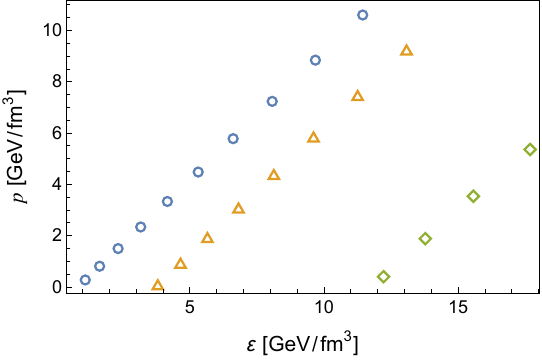}
      \end{minipage} 
      \begin{minipage}[h]{0.33\columnwidth}
        \centering
        \includegraphics[width=0.9\columnwidth]{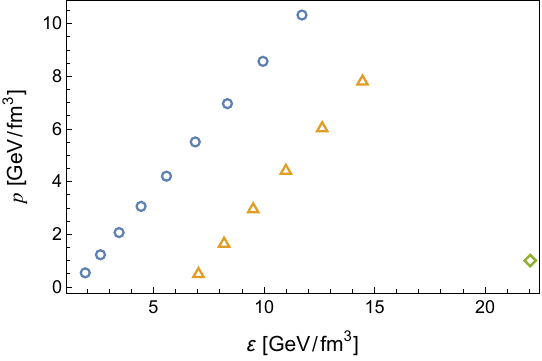}
      \end{minipage}    
      \\\\
      \begin{minipage}[h]{0.33\columnwidth}
        \centering
        \includegraphics[width=0.9\columnwidth]{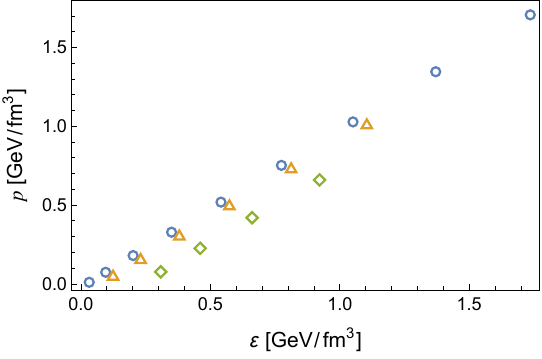}
      \end{minipage} 
      \begin{minipage}[h]{0.33\columnwidth}
        \centering
        \includegraphics[width=0.9\columnwidth]{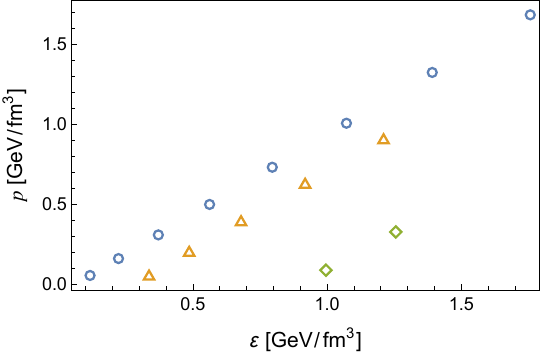}
      \end{minipage} 
      \begin{minipage}[h]{0.33\columnwidth}
        \centering
        \includegraphics[width=0.9\columnwidth]{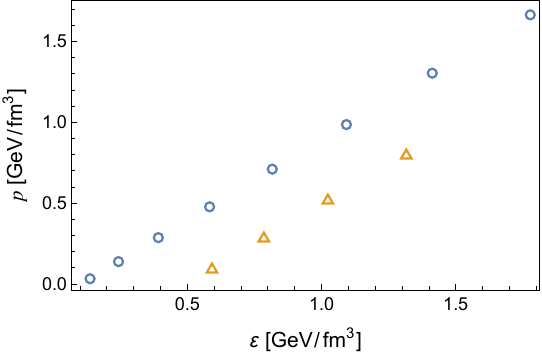}
      \end{minipage}    
      \end{tabular}
       \caption{The equation of state at $L^{-1}=323\ \mathrm{MeV}$ (upper) and $L^{-1}=170\ \mathrm{MeV}$ (lower) for $k_2=5$ (left), $k_2=20$ (middle), and $k_2=35$ (right).
       Blue circles, orange triangles and green diamonds are for $B = 0.4/L$, $0.6/L$ and $0.8/L$, respectively.}
        \label{fig:eos}
  \end{figure}

We plot the equations of state (EoSs) in the baryonic phase 
in Fig.~\ref{fig:eos}.
The pressure $p$ is given by $-\Omega/V$.
The energy density $\varepsilon$ is equal to the free energy density
because we work in the zero temperature limit,
\begin{align}
  \varepsilon=\frac{F}{V}=\frac{\Omega}{V}+\mu_Bd_B \ .\nn
\end{align}
The EoS $p=p(\varepsilon)$ is approximated by a linear function with the gradient nearly equal to $1$.
A closer inspection shows that the EoS is best fitted by a quadratic function of $\varepsilon$ with a tiny coefficient of $\varepsilon^2$.
{
For $L^{-1}=323 \mathrm{MeV}$,

$k_2=5$
\begin{equation*}
p(\varepsilon)=
\left\{
\begin{aligned}
	-0.19+1.0 \varepsilon-0.000069 \varepsilon^2
	\quad (B=0.4/L)
	\\
	-0.93+1.0 \varepsilon-0.0012 \varepsilon^2
	\quad (B=0.6/L)
	\\
	-3.1+1.065 \varepsilon-0.0073 \varepsilon^2
	\quad (B=0.8/L)
\end{aligned}
\right.
\end{equation*}

$k_2=20$
\begin{equation*}
        p(\varepsilon)=
        \left\{
        \begin{aligned}
        -0.73+1.0 \varepsilon-0.000073 \varepsilon^2
	\quad (B=0.4/L)
        \\
        -3.7+1.0 \varepsilon-0.0014 \varepsilon^2
	\quad (B=0.6/L)
        \\
        -13+1.3 \varepsilon-0.013 \varepsilon^2
	\quad (B=0.8/L)
        \end{aligned}
        \right.
\end{equation*}

$k_2=35$
\begin{align*}
        p(\varepsilon)=
        \left\{
        \begin{aligned}
	-1.3+1.0 \varepsilon-0.000072 \varepsilon^2
	\quad (B=0.4/L)
	\\
	-6.5+1.0 \varepsilon-0.0017 \varepsilon^2
	\quad (B=0.6/L)
	\end{aligned}
	\right.
\end{align*}

and for $L^{-1}=170 \mathrm{MeV}$,

$k_2=5$
\begin{align*}
        p(\varepsilon)=
        \left\{
        \begin{aligned}
	-0.022+1.0 \varepsilon-0.0035 \varepsilon^2
	\quad (B=0.4/L)
	\\
	-0.081+1.0 \varepsilon-0.029 \varepsilon^2
	\quad (B=0.6/L)
	\\
	-0.24+1.1 \varepsilon-0.081 \varepsilon^2
	\quad (B=0.8/L)
	\end{aligned}
	\right.
\end{align*}

$k_2=20$
\begin{align*}
        p(\varepsilon)=
        \left\{
        \begin{aligned}
	-0.065+1.0 \varepsilon-0.0038 \varepsilon^2
	\quad (B=0.4/L)
	\\
	-0.30+1.03101 \varepsilon-0.034 \varepsilon^2
	\quad (B=0.6/L)
	\\
	-0.36+0.071 \varepsilon+0.38 \varepsilon^2
	\quad (B=0.8/L)
	\end{aligned}
	\right.
\end{align*}

$k_2=35$
\begin{align*}
        p(\varepsilon)=
        \left\{
        \begin{aligned}
	-0.11+1.0 \varepsilon-0.0038 \varepsilon^2
	\quad (B=0.4/L)
	\\
	-0.52+1.1 \varepsilon-0.042 \varepsilon^2
	\quad (B=0.6/L).
	\end{aligned}
	\right.
\end{align*}
}

\begin{figure}[htbp]
    \begin{tabular}{cc}
      \begin{minipage}[h]{0.33\columnwidth}
        \centering
        \includegraphics[width=0.9\columnwidth]{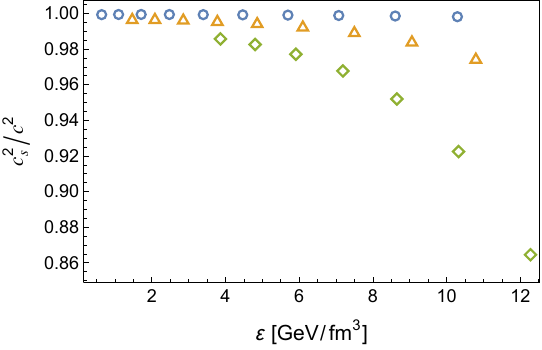}
      \end{minipage} 
      \begin{minipage}[h]{0.33\columnwidth}
        \centering
        \includegraphics[width=0.9\columnwidth]{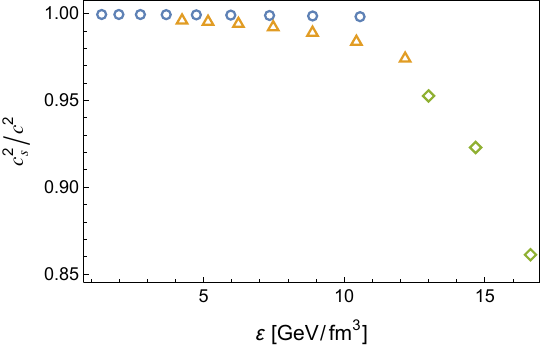}
      \end{minipage} 
      \begin{minipage}[h]{0.33\columnwidth}
        \centering
        \includegraphics[width=0.9\columnwidth]{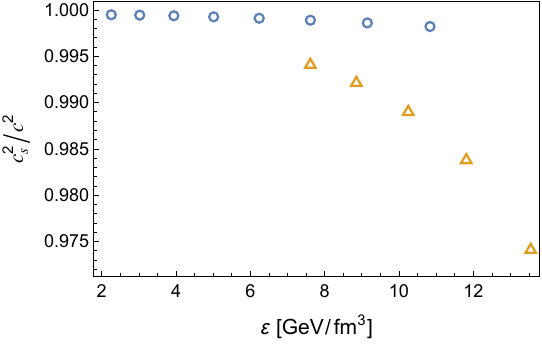}
      \end{minipage}    
      \\\\
      \begin{minipage}[h]{0.33\columnwidth}
        \centering
        \includegraphics[width=0.9\columnwidth]{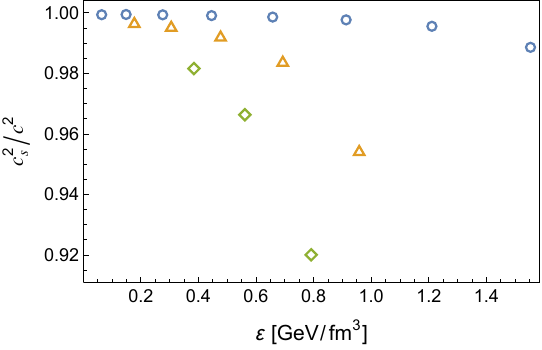}
      \end{minipage} 
      \begin{minipage}[h]{0.33\columnwidth}
        \centering
        \includegraphics[width=0.9\columnwidth]{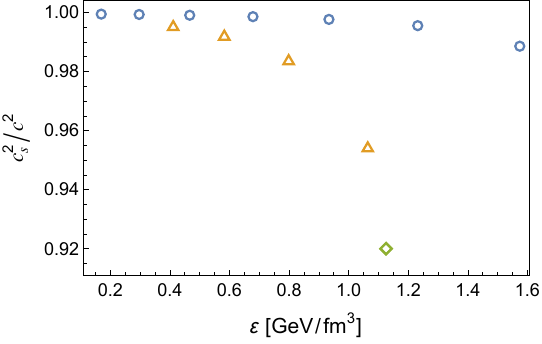}
      \end{minipage} 
      \begin{minipage}[h]{0.33\columnwidth}
        \centering
        \includegraphics[width=0.9\columnwidth]{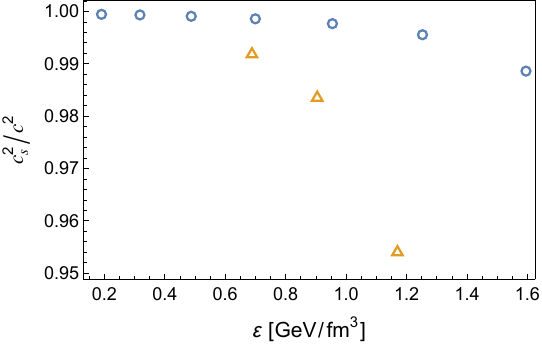}
      \end{minipage}    
      \end{tabular}
       \caption{The speed of sound normalized by the speed of light as a function of the energy density at $L^{-1}=323\ \mathrm{MeV}$ (upper) and $L^{-1}=170\ \mathrm{MeV}$ (lower) for $k_2=5$ (left), $k_2=20$ (middle), and $k_2=35$ (right).
       Blue circles, orange triangles and green diamonds are for $B = 0.4/L$, $0.6/L$ and $0.8/L$, respectively.}
        \label{fig:sv}
  \end{figure}
\begin{figure}[h]
	\centering
	\includegraphics[width=7.5cm]{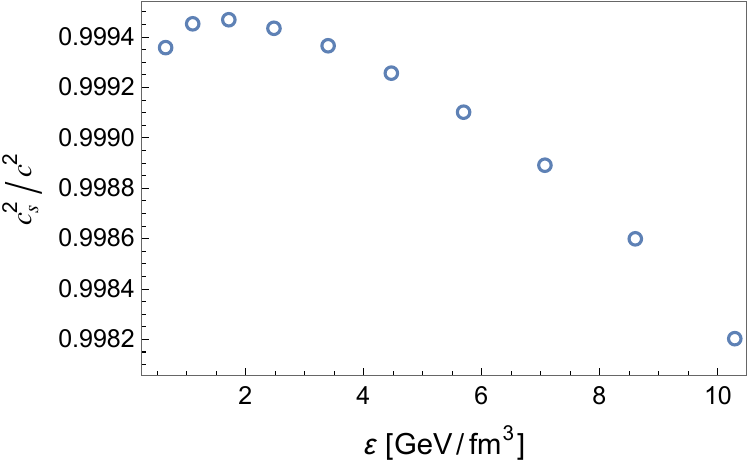}
	\caption{A gradual peak is shown in the speed of sound for $L^{-1}=323\ \mathrm{MeV},\ k_2=5,\ B=0.4/L$.
	\label{fig:peak}}
\end{figure}

Once the EoS is obtained,
we find the speed of sound $c_s$ in the baryonic phase by using the
formula  $c_s^2=\frac{\partial p}{\partial \varepsilon}$.
We show the results in Fig.~\ref{fig:sv} with 
$c$ being the speed of light.
For all the cases, the speed of sound is
nearly equal to that of light.
It does not approach the conformal limit $c_s^2=1/3$.
This is because the present holographic model of QCD
is not applicable in the conformal regime, which
is realized in the high-density and the weak coupling region of QCD.
We note that a peak structure in the speed of sound emerges for several choices of the parameters.
As an example, we show the case of $L^{-1}=323\ \mathrm{MeV},\ k_2=5,\ B=0.4/L$ in Fig.~\ref{fig:peak}.
However, the magnitude of the peak is gradual.

\begin{figure}[htbp]
    \begin{tabular}{cc}
      \begin{minipage}[h]{0.33\columnwidth}
        \centering
        \includegraphics[width=0.9\columnwidth]{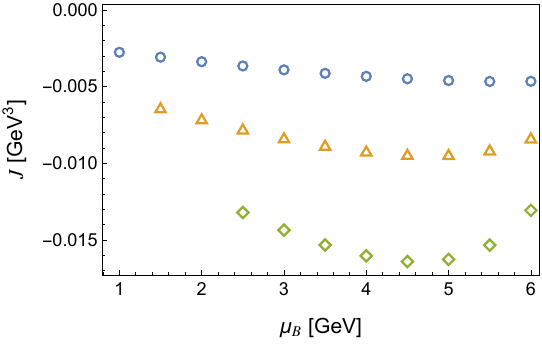}
      \end{minipage} 
      \begin{minipage}[h]{0.33\columnwidth}
        \centering
        \includegraphics[width=0.9\columnwidth]{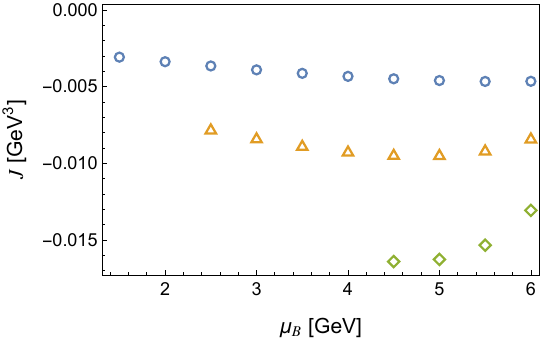}
      \end{minipage} 
      \begin{minipage}[h]{0.33\columnwidth}
        \centering
        \includegraphics[width=0.9\columnwidth]{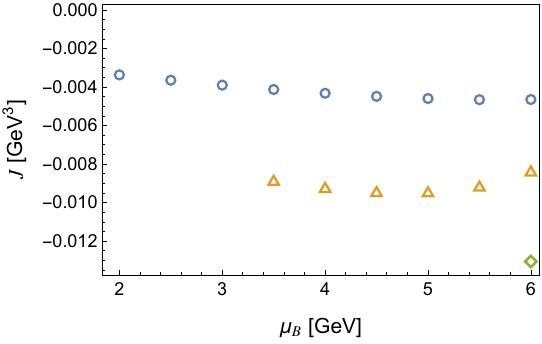}
      \end{minipage}    
      \\\\
      \begin{minipage}[h]{0.33\columnwidth}
        \centering
        \includegraphics[width=0.9\columnwidth]{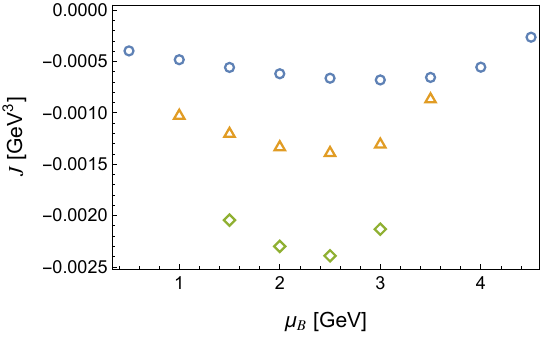}
      \end{minipage} 
      \begin{minipage}[h]{0.33\columnwidth}
        \centering
        \includegraphics[width=0.9\columnwidth]{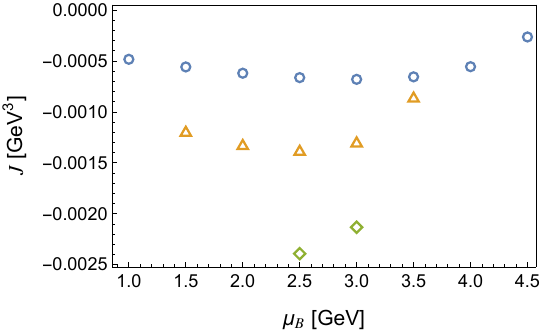}
      \end{minipage} 
      \begin{minipage}[h]{0.33\columnwidth}
        \centering
        \includegraphics[width=0.9\columnwidth]{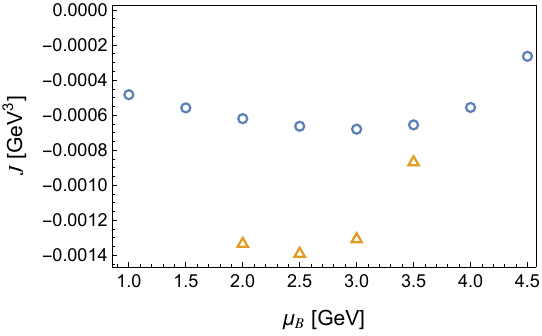}
      \end{minipage}    
      \end{tabular}
       \caption{The spatial component of axial-isovector meson condensate as a function of $\mu_B$ at $L^{-1}=170\ \mathrm{MeV}$ (upper) and $L^{-1}=170\ \mathrm{MeV}$ (lower) for $k_2=5$ (left), $k_2=20$ (middle), and $k_2=35$ (right).
       Blue circles, orange triangles, and green diamonds are for $B = 0.4/L$, $0.6/L$, and $0.8/L$, respectively.}
        \label{fig:j}
  \end{figure}

Finally, the axial-isovector condensate $J$ in the baryonic phase is 
plotted in Fig.~\ref{fig:j}.
The GKP-W relation (\ref{gkpw:J}) suggests that $J$ is interpreted
as the condensate of axial-isovector mesons, which arise
from Kaluza-Klein reduction of 5D axial-isovector gauge fields
along the $z$ direction.\footnote{
In the holographic model of QCD on the basis of D4/D8-branes,
it is shown \cite{Hashimoto:2008zw}
that the axial flavor current 
is written in terms of the sum of the gradient of the massless pion field and
an infinite number of axial-vector meson fields.}
For recent works on axial-isovector meson condensates in high-density QCD, see 
\cite{Suenaga:2023xwa,Brandt:2018bwq,Adhikari:2020qda,Brauner:2016lkh,Chen:2024cxh}.

\subsection{M-R plot}

\begin{figure}[htbp]
    \begin{tabular}{cc}
      \begin{minipage}[h]{0.33\columnwidth}
        \centering
        \includegraphics[width=0.9\columnwidth]{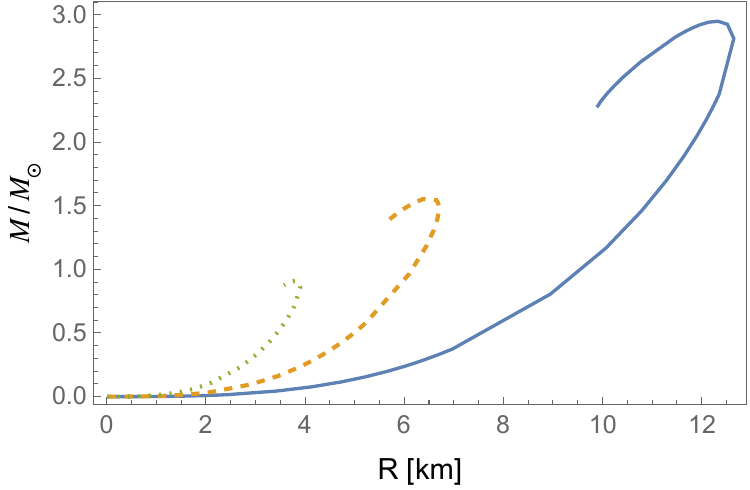}
      \end{minipage} 
      \begin{minipage}[h]{0.33\columnwidth}
        \centering
        \includegraphics[width=0.9\columnwidth]{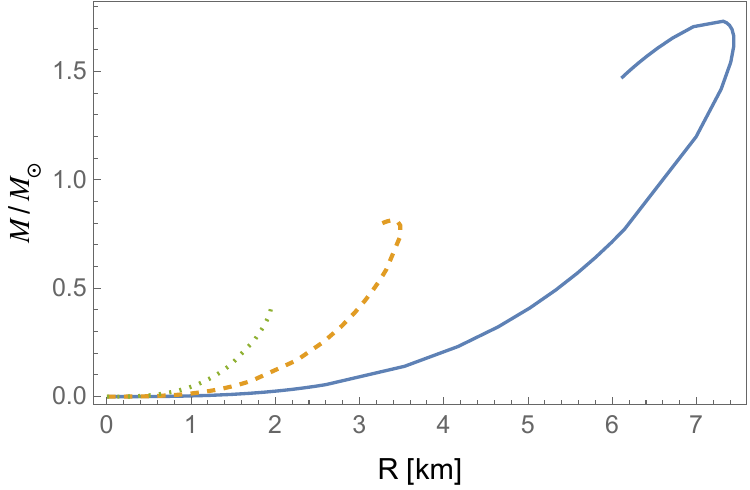}
      \end{minipage} 
      \begin{minipage}[h]{0.33\columnwidth}
        \centering
        \includegraphics[width=0.9\columnwidth]{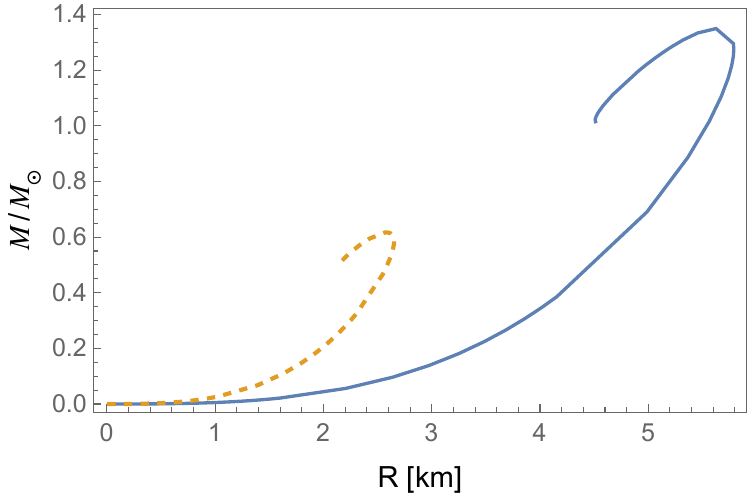}
      \end{minipage}    
      \\\\
      \begin{minipage}[h]{0.33\columnwidth}
        \centering
        \includegraphics[width=0.9\columnwidth]{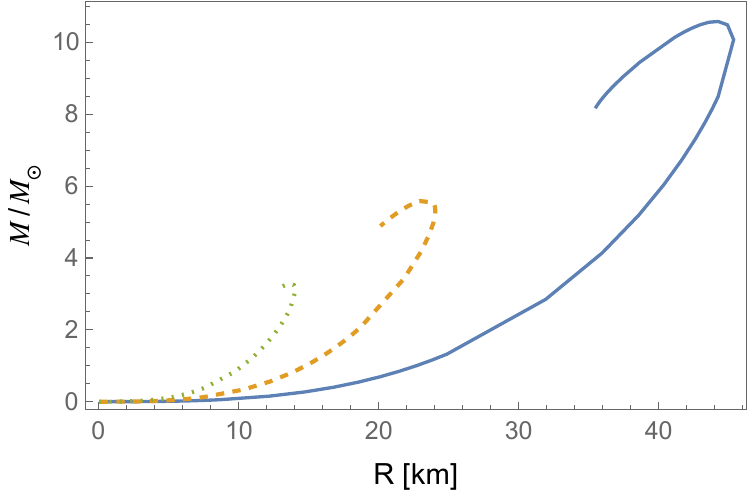}
      \end{minipage} 
      \begin{minipage}[h]{0.33\columnwidth}
        \centering
        \includegraphics[width=0.9\columnwidth]{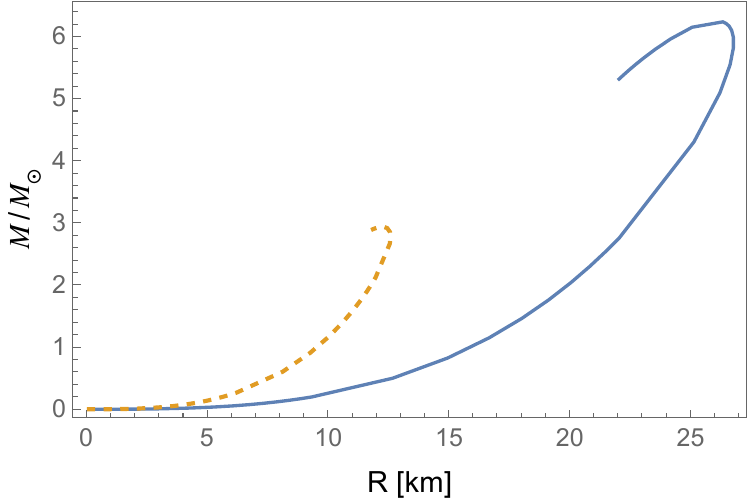}
      \end{minipage} 
      \begin{minipage}[h]{0.33\columnwidth}
        \centering
        \includegraphics[width=0.9\columnwidth]{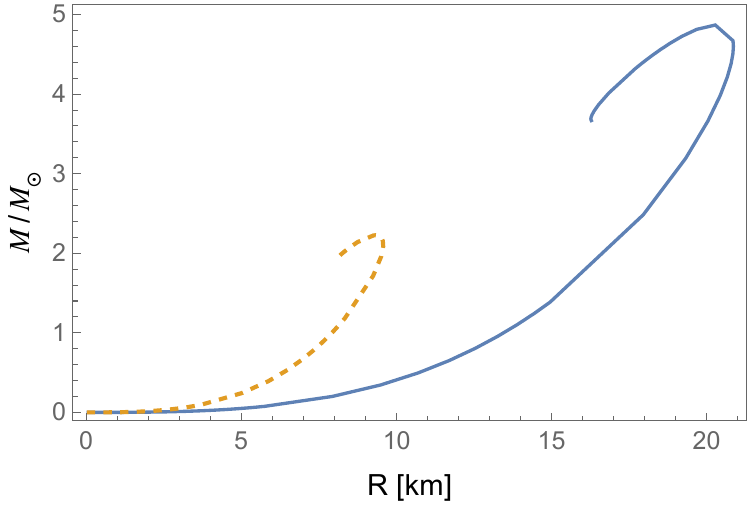}
      \end{minipage}    
      \end{tabular}
       \caption{{M-R plots resulting from the EoSs for } $L^{-1}=323\ \mathrm{MeV}$ (upper) and $L^{-1}=170\ \mathrm{MeV}$ (lower) for $k_2=5$ (left), $k_2=20$ (middle), and $k_2=35$ (right).
       The blue solid line, orange dashed line, and green dotted line are for $B=0.4/L$, $0.6/L$, and $0.8/L$, respectively. The neutron star mass is measured in units of the solar mass $M_\odot$.}
        \label{fig:mr}
  \end{figure}

The TOV equation \cite{Oppenheimer:1939ne}
reads 
\begin{equation}
	\frac{dp(r)}{dr}=-\frac{G\varepsilon(r)M(r)}{r^2}
	\left[1+\frac{p(r)}{\varepsilon(r)}\right]
	\left[1+\frac{4\pi r^3 p(r)}{M(r)}\right]
	\left[1-\frac{2GM(r)}{r}\right]^{-1}\ .
\label{tov}
\end{equation}
Here, $G$ is the Newton constant, 
$r$ is the radial coordinate of a neutron star, and
$M(r)$ is the energy of the neutron star that is stored inside
the region up to $r$, which obeys
\begin{equation}
	\frac{d M(r)}{dr}=4\pi r^2\varepsilon(r) \ .
\label{eneeqn}
\end{equation}
We solve these equations 
together with the EoS $p=p(\varepsilon)$ with
the boundary conditions $p(r=0)=p_0$ and $M(r=0)=0$.
The mass of a neutron star is given by $M(r=R)$, with $R$ 
being the radius of the neutron star, which is defined by the smallest
value of $R$ satisfying $p(R)=0$.
The M-R plot is obtained by depicting a curve that interpolates
the set of the mass $M(R)$ and the radius $R$, which is computed for different initial values of $p_0$.
The results are shown in Fig.~\ref{fig:mr}.

\section{Discussion}

\begin{figure}[htbp]
    \begin{tabular}{cc}
      \begin{minipage}[h]{0.33\columnwidth}
        \centering
        \includegraphics[width=0.9\columnwidth]{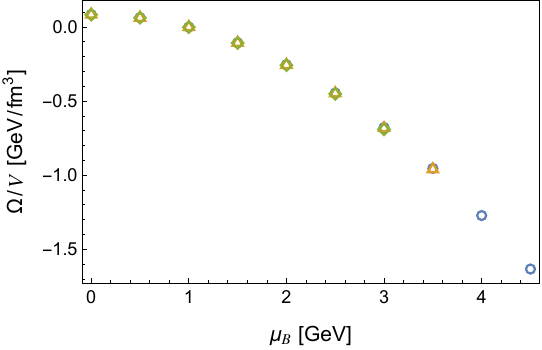}
      \end{minipage} 
      \begin{minipage}[h]{0.33\columnwidth}
        \centering
        \includegraphics[width=0.9\columnwidth]{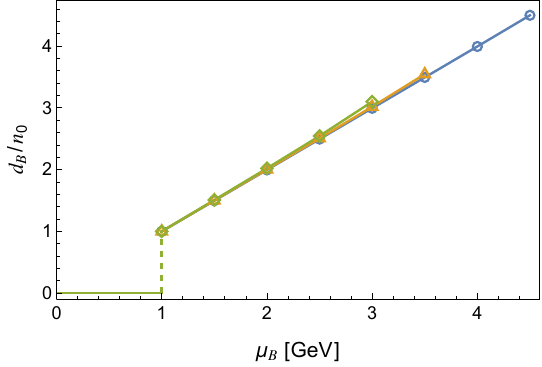}
      \end{minipage} 
      \begin{minipage}[h]{0.33\columnwidth}
        \centering
        \includegraphics[width=0.9\columnwidth]{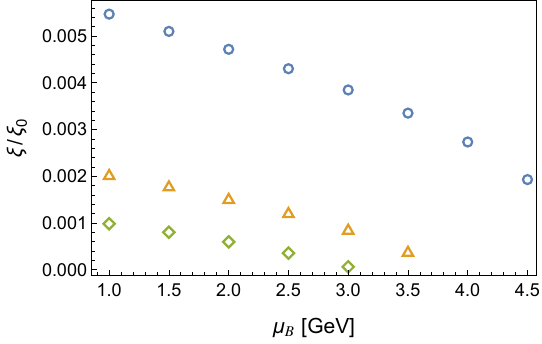}
      \end{minipage}    
      \\\\
      \begin{minipage}[h]{0.33\columnwidth}
        \centering
        \includegraphics[width=0.9\columnwidth]{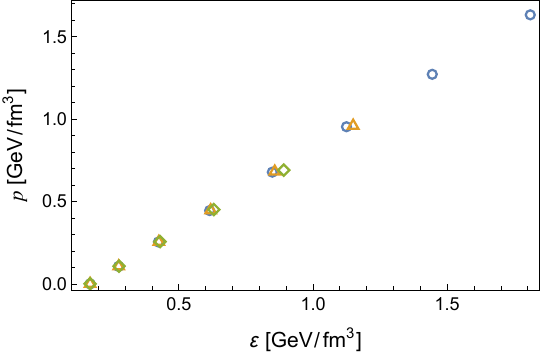}
      \end{minipage} 
      \begin{minipage}[h]{0.33\columnwidth}
        \centering
        \includegraphics[width=0.9\columnwidth]{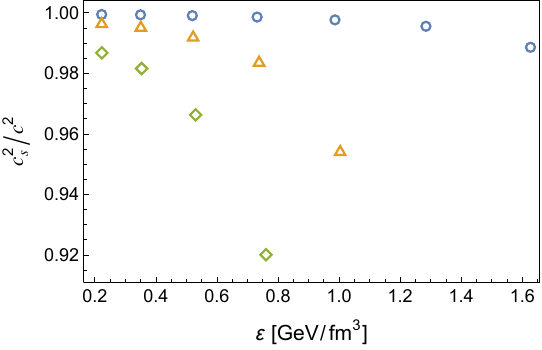}
      \end{minipage} 
      \begin{minipage}[h]{0.33\columnwidth}
        \centering
        \includegraphics[width=0.9\columnwidth]{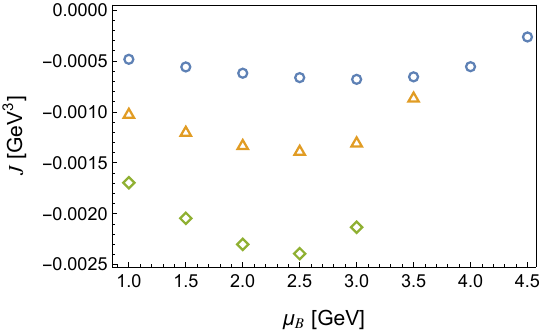}
      \end{minipage}    
      \end{tabular}
       \caption{The properties for $L^{-1}=170\ \mathrm{MeV}$ with
       $(k_2,B)=(57.9,0.4/L)$ (blue circles), $(k_2,B)=(11.4,0.6/L)$ (orange triangles), and $(k_2,B)=(3.6,0.8/L)$ (green diamonds).
       The grand potential density (upper left), the baryon number density (upper middle), the chiral condensate in the baryonic phase (upper right), the EoS in the baryonic phase (lower left), the speed of sound in the baryonic phase (lower middle), and the condensate of the spatial component of the axial-isovector meson in the baryonic phase (lower right). 
       Note that lots of three colors of marks overlap with each other in the plots of ground potential, baryon number density, and pressure.}
        \label{fig:fixingdB}
  \end{figure}

\begin{figure}[h]
	\centering
	\includegraphics[width=10cm]{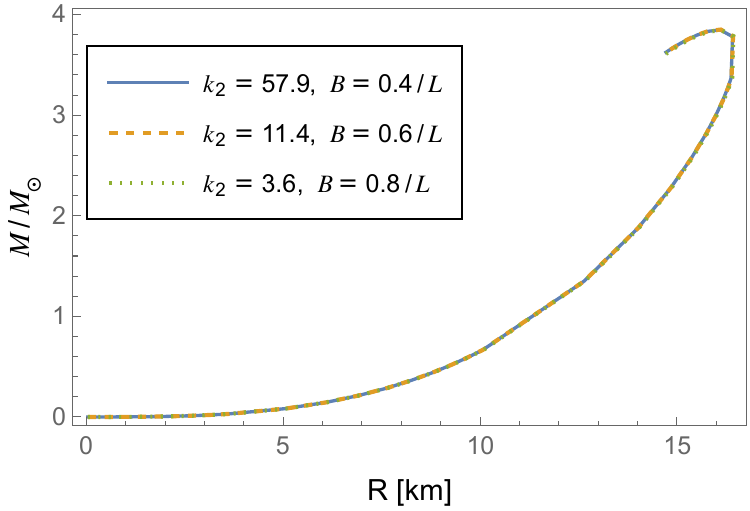}
	\caption{M-R plots {in $L^{-1}=170\ \mathrm{MeV}$} for $(k_2,B)=(57.9,0.4/L)$ (blue solid line), $(k_2,B)=(11.4,0.6/L)$ (orange dashed line), and $(k_2,B)=(3.6,0.8/L)$ (green dotted line).
	\label{fig:fixingmr}}
\end{figure}

In this paper, we have analyzed numerically the high-density QCD
phase by making a sample choice of $k_2$ and $B$.
It is found that the critical values of baryon number chemical potential $\mu_B$ and baryon number density $d_B$ depend on
how to choose them. Naively, it is expected that equating the critical
values with input data fixes $k_2$ and $B$ uniquely. 
This is not the case, however.
To see this,
we assume that
the phase transition from the nonbaryonic phase to the baryonic phase occurs
at $\mu_B=1\ \mathrm{GeV}$ with
the critical baryon number density given by $d_B=n_0=0.17\ \mathrm{fm^{-3}}$,
as an example. 
Figure~\ref{fig:fixingdB} shows that three different choices of $k_2$ and $B$
lead to the same critical values of $\mu_B$ and $d_B$.
It is also seen that, as shown in Fig.~\ref{fig:fixingmr}, the M-R plot
does not depend on the choice of the parameters either.
In order to fix $k_2$ and $B$ uniquely, an input about the critical values of
the chiral condensate, the speed of sound, or the spatial component of the axial-isovector meson condensate is necessary.
We leave a detailed study in this line for future work.

\begin{figure}[h]
	\centering
	\includegraphics[width=10cm]{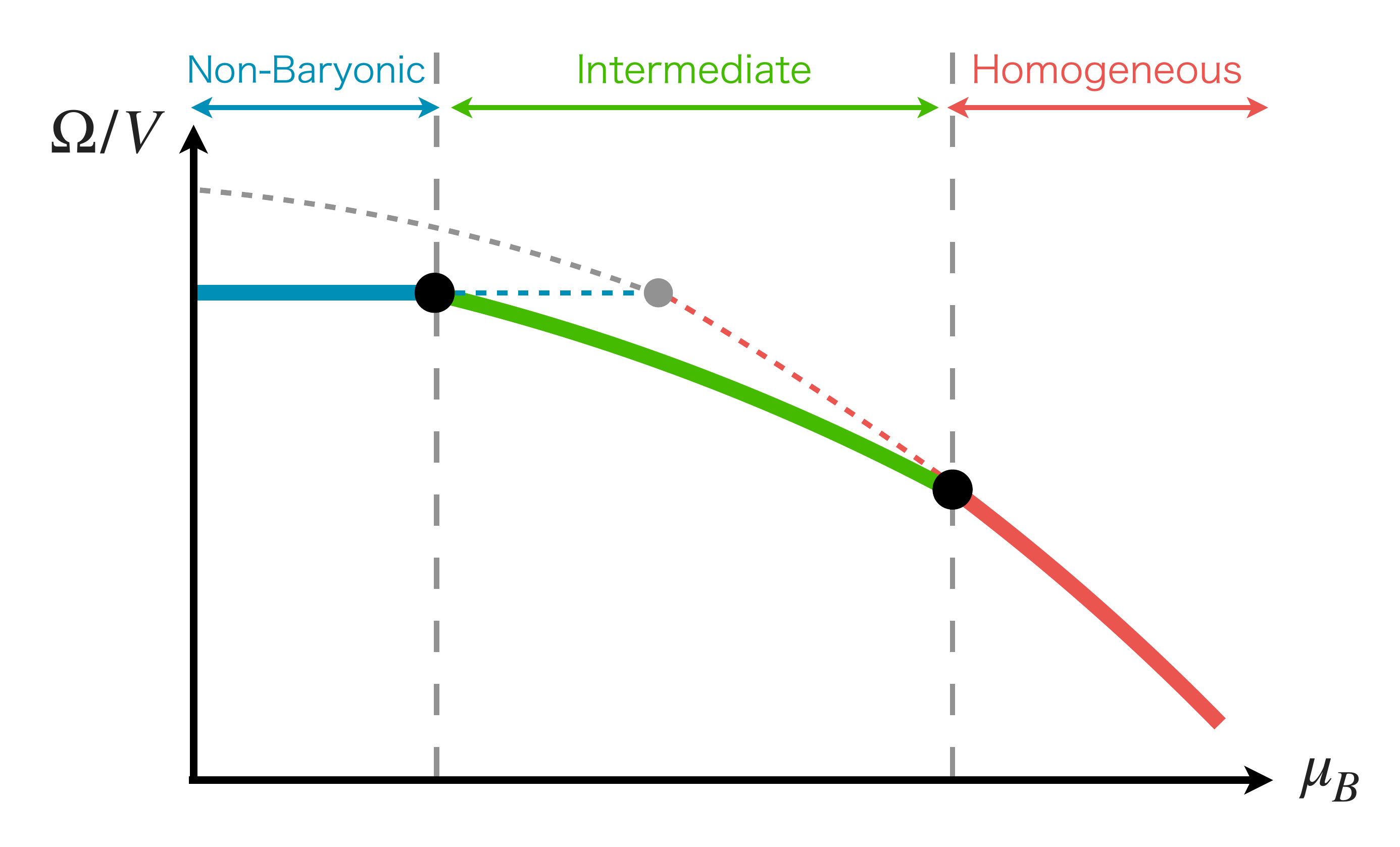}
	\vspace{-0.5cm}
	\caption{{A {schematic} picture of the intermediate phase.}
	\label{fig:intermediate}}
\end{figure}

We have assumed that, as $\mu_B$ grows, the nonbaryonic
phase makes a phase transition to the baryonic phase, which
is described by the homogeneous ansatz.
It is found that the theoretical prediction from the baryonic phase 
reproduces the properties of the baryonic matter phase very well when making an
appropriate choice of the parameters in the model.
For $L^{-1}=323\ \mathrm{MeV}$, however, the critical values of
$\mu_B$ and $d_B$ are larger than what is naturally expected from
the viewpoint of hadron physics. This implies the existence of an
intermediate phase between the nonbaryonic phase and the baryonic phase.
See Fig. \ref{fig:intermediate} for a typical phase diagram
that implements this idea.
Such a phase must incorporate the many-body effects
of nucleons that cannot be captured by the homogeneous ansatz, because 
this ansatz should be regarded as an effective description of the nuclear
matter in terms of a mean field.
So far, there have been many attempts toward a deeper understanding of
the baryonic matter as the many-body system of nucleons.
One of them is
the Skyrmion or the instanton crystal, see, for instance,
\cite{Klebanov:1985qi,Nawa:2008uv, Jarvinen:2020xjh}.

Here, we show that the magnitude of the axial vector meson condensate $J$ found in the previous section is consistent with an estimate in the Skyrme model.
The pion field in the Skyrme model is described by the hedgehog ansatz \cite{Pauli:1942kwa,Skyrme:1961vq,Skyrme:1962vh} 
\[
	U(\vec{r}) = e^{i \vec{\tau} \cdot \vec{\pi} / f_\pi} = e^{i\vec{\tau} \cdot \hat{r} F(r)},
\]
where $\hat{r}$ is the unit vector of the three-dimensional polar coordinate and $f_\pi=64.5\ \mathrm{MeV}$ is the pion decay constant.
From the hedgehog ansatz, the pion field reads
\[
	\pi^a = f_\pi \hat{r}^a F(r).
\]
By noting that the axial current $A_\mu$ is described by $A_\mu \sim -f_\pi \partial_\mu \pi$, we obtain
\[
	A_i^a 
	\sim
	-f_\pi^2 \partial_i (\hat{r}^a F(r))
	=
	 - f_\pi^2
	\left[
		F(r) \frac{\delta^a_i - \hat{r}^a \hat{r}_i}{r}
		+\hat{r}^a \hat{r}_i \partial_r F(r)
	\right].
\]
Averaging the axial current over the angular directions gives
\[
	\bar{A}_i^a \sim -f_\pi^2 
	\left[
		\frac{2}{3r} F(r) + \frac{1}{3} \partial_r F(r)
	\right]
	\delta^a_i.
\]

We further take an average of the axial current along the radial direction. 
To this end, we compute
\begin{align*}
	\int_0^{r_*} dr\, r^2 \bar{A}_i^a
	\sim &
	-f_\pi^2\, \delta_i^a 
	\int_0^{r_*} dr\, r^2
	\left[
		\frac{2}{3r}F(r)
		+\frac{1}{3} \partial_r F(r)
	\right]
	\\=&
	-\frac{f_\pi^2}{3} \, \delta_i^a
	\left[
		r^2 F(r)
	\right]_0^{r_*}
	\\=&
	-\frac{f_\pi^2}{3} \, \delta_i^a
	r_*^2 F(\tilde{r}_*),
\end{align*}
where $\tilde{r}_*=ef_\pi r_*$, and $e=5.45$ is the coupling constant in the Skyrme term.
Note that we have used $F(0)=\mathrm{finite}$.
The average of the axial current over a ball of size $r_*$ is given by
\begin{align*}
	\langle \bar{A}_i^a \rangle_{r_*}
	\equiv&
	\frac{1}{r_*^3/3} \int_0^{r_*} dr\, r^2 \bar{A}_i^a 
	\\=&
	-e f_\pi^3\, \delta_i^a \frac{1}{\tilde{r}_*} F(\tilde{r}_*).
\end{align*}
Assuming $\tilde{r}_*=\mathcal{O}(1)$, the value of $\langle \bar{A}_i^a \rangle_{r_*}$ is evaluated as
\begin{equation*}
	\langle \bar{A}_i^a \rangle_{r_*}
	\sim
	- e f_\pi^3
	= -5.45 \times (64.5\ \mathrm{MeV})^3
	\simeq -1.5\times 10^{-3}\ \mathrm{GeV}^3.
\end{equation*}
This result is consistent with the order of $J$ in our results.

The holographic model of QCD in this paper allows us to turn on
the axial-isovector chemical potential $\hat{\phi}$.
Although the physical meaning of it is unclear to us,
it would be interesting to study what the QCD phase diagram looks
like when extending it to the $\hat{\phi}$ direction.

An extension to the three-flavor case is an attractive future work.
One of the motivations for this is to discuss if and how
the hyperon puzzle is resolved in the three-flavor holographic model~\cite{Demorest:2010bx}.
The hyperon puzzle states that an incorporation of the
strange quarks into the baryonic matter softens the EoS so much
that no neutron stars are allowed to exist whose mass is larger
than $2M_\odot$.
As seen above, the EoSs obtained in this paper tend to be stiff.
This might imply that even in the presence of the
strange quark, the EoS remains stiff enough to lead to 
neutron stars whose mass is consistent with the astronomical observations.

\section*{Acknowledgments}
We thank Sota Akahane, Masayuki Asakawa, Lorenzo Bartolini, Kenji Fukushima, Sven Bjarke Gudnason, Noriyoshi Ishii, Daisuke Jido, Toru Kojo, Shin Nakamura, Shigeki Sugimoto, Hajime Togashi, and Naoki Yamamoto for their valuable discussions.

This work was supported by JSPS KAKENHI Grant Numbers JP24KJ16200 and JP24K17054 and JST SPRING Grant Number JPMJSP2138.

\appendix

\section{Hamilton-Jacobi equation and boundary conditions}
\label{HJeqn}
Start with the action $S=S_{\rm{bulk}}+S_{\rm{bdry}}$ with
\begin{align}
  S_{\rm{bulk}}&=\int_{z_0}^{z_1}dz L(q,q^{\prime})
=\int_{z_0}^{z_1}dz\, \left(pq^{\prime}-H(q,p)\right) \ ,
\\
S_{\rm{bdry}}&=L_1(q(z_1))-L_0(q(z_0)) \ .
\end{align}
Here, $p=\partial L/\partial q^{\prime}$.
The variation of the action is
\begin{align}
  \delta S=
\left(p(z_1)+\frac{\partial L_1}{\partial q(z_1)}\right)
\delta q(z_1)
-
\left(p(z_0)+\frac{\partial L_0}{\partial q(z_0)}\right)
\delta q(z_0)
+\int dz \left({\rm{bulk~terms}}\right) \ .
\end{align}
The boundary condition is given by requiring that the boundary term
vanish,
\begin{align}
\left(p(z_1)+\frac{\partial L_1}{\partial q(z_1)}\right)
\delta q(z_1) = 0 \ ,
~~\mbox{and}~~
\left(p(z_0)+\frac{\partial L_0}{\partial q(z_0)}\right)
\delta q(z_0) = 0 \ .
\end{align}
These lead to either the Dirichlet or the Neumann boundary conditions.

We first solve the equation of motion for $q$ under the condition
\begin{align}
  q(z_1)=q_1 \ ,~~~q(z_0)=q_0 \ .
\label{q0q1}
\end{align}
Inserting the solution into $S$ gives the on-shell action as a 
function of $(q_1,z_1)$ and $(q_1,z_1)$,
\begin{align}
  \ol S(q_0,z_0;q_1,z_1)\equiv
S_{\rm{bulk}}(q,q^{\prime})\big|_{\rm{EoM}}+L_1(q_1)-L_0(q_0) \ .
\end{align}
The on-shell bulk action obeys the Hamilton-Jacobi equation
\begin{align}
\delta  S_{\rm{bulk}}(q,q^{\prime})\big|_{\rm{EoM}}
=p(z_1)\delta q_1-H(q_1,p(z_1))\delta z_1
-p(z_0)\delta q_0+H(q_0,p(z_0))\delta z_0 \ .
\end{align}
It follows that
\begin{align}
\delta \ol S(q_0,z_0;q_1,z_1)=
\left(p(z_1)+\frac{\partial L_1}{\partial q_1}\right)
\delta q_1
-
\left(p(z_0)+\frac{\partial L_0}{\partial q_0}\right)
\delta q_0
-H(q_1,p(z_1))\delta z_1
+H(q_0,p(z_0))\delta z_0 \ .
\end{align}
In an application of this result to holographic renormalization,
we set
\begin{align}
  \delta q_0=\delta (\mbox{UV boundary values}) \ ,~~~
\delta q_1=0 \ ,~~~
z_0,z_1=\mbox{fixed}  \ .
\end{align}
We impose the second condition because the IR boundary value $q_1$
is independent of $q_0$. We find that the variation of the on-shell
action receives the contributions only from the UV boundary.

Instead of (\ref{q0q1}), we may impose the Neumann boundary condition
at the IR boundary $z=z_1$. Then, $q_1$ is fixed to be
$q_1=q_1^{\ast}(q_0,z_0,z_1)$ by solving the Hamilton-Jacobi equation
\begin{align}
  \frac{\partial\ol S}{\partial q_1}
=
p(z_1)+\frac{\partial L_1}{\partial q_1}=0 \ .
\label{N:delq1}
\end{align}
Inserting this solution into $\ol S$ leads to another on-shell action
\begin{align}
  \ol S^{\prime}(q_0,z_0;z_1)
\equiv\ol S(q_0,z_0;q_1=q_1^{\ast}(q_0,z_0,z_1),z_1) \ .
\end{align}
It is seen that the resultant on-shell action obeys the Hamilton-Jacobi equation.
For instance,
\begin{align}
  \frac{\partial\ol S^{\prime}}{\partial q_0}
=
\left(\frac{\partial \ol S}{\partial q_0}\right)_{q_1,z_0,z_1}
(q_1=q_1^\ast)
+
\left(\frac{\partial \ol S}{\partial q_1}\right)_{q_0,z_0,z_1}
(q_1=q_1^\ast)\cdot \frac{\partial q_1^{\ast}}{\partial q_0}
=
-\left(p(z_0)+\frac{\partial L_0}{\partial q_0}\right)
\Bigg|_{q_1=q_1^{\ast}}\ .
\end{align}
Here, (\ref{N:delq1}) and the Hamilton-Jacobi equation for $\ol S$ are used.
A similar computation gives
\begin{align}
  \frac{\partial\ol S^{\prime}}{\partial z_0}
=H(q_0,p(z_0))\big|_{q_1=q_1^{\ast}} \ ,~~~
  \frac{\partial\ol S^{\prime}}{\partial z_1}
=-H(q_1,p(z_1))\big|_{q_1=q_1^{\ast}} \ .
\end{align}
In an application of this result to holographic renormalization,
we set
\begin{align}
  \delta q_0=\delta (\mbox{UV boundary values}) \ ,
~~~z_0,z_1=\mbox{fixed}  \ .
\end{align}
Again, the variation of the on-shell
action receives the contributions only from the UV boundary.

\bibliography{refs}

\end{document}